\documentclass[12pt]{article}
\pdfoutput=1
\usepackage{putex}
\usepackage{feyn}
\usepackage[vcentermath]{youngtab}
\usepackage{slashed}

\usepackage{graphicx}
\usepackage{epstopdf}
\usepackage{enumerate}
\usepackage{cite}
\usepackage{tensor}
\usepackage{slashed}
\usepackage{amsmath}
\usepackage{amssymb}
\usepackage{mathrsfs}
\usepackage{comment}

\usepackage{hyperref}

\numberwithin{equation}{section}

\newcommand {\be} {\begin {equation}}
\newcommand {\ee} {\end {equation}}

\newcommand {\bes} {\begin {equation*}}
\newcommand {\ees} {\end {equation*}}

\newcommand{\eps}{\epsilon}

\newcommand{\deh}{\hat{\partial}}

\newcommand{\beq}{\begin{equation}}
\newcommand{\eeq}{\end{equation}}

\def\be{ \begin{equation} }
\def\ee{ \end{equation} }

\begin{document}

\preprint{PUPT-2495}

\institution{PU}{Department of Physics, Princeton University, Princeton, NJ 08544}

\title{
Anomalous dimensions in CFT with weakly broken higher spin symmetry
}

\authors{Simone Giombi\worksat{\PU} and Vladimir Kirilin\worksat{\PU}
}

\abstract{In a conformal field theory with weakly broken higher spin symmetry, the leading order anomalous dimensions of the broken 
currents can be efficiently determined from the structure of the classical non-conservation equations. We apply this method 
to the explicit example of $O(N)$ invariant scalar field theories in various dimensions, including the large $N$ critical $O(N)$ model in general $d$, the Wilson-Fisher fixed point in $d=4-\eps$, cubic scalar models in $d=6-\eps$ and the nonlinear sigma model in $d=2+\eps$. Using
information from the $d=4-\eps$ and $d=2+\eps$ expansions, we obtain some estimates for the dimensions of the higher spin 
operators in the critical 3d $O(N)$ models for a few low values of $N$ and spin.}

\date{}
\maketitle

\tableofcontents

\section{Introduction and Summary}
\label{intro}

The spectrum of a $d$-dimensional conformal field theory \cite{Polyakov:1970xd} consists of local operators labelled by conformal dimension $\Delta$, 
a representation $R$ of $SO(d)$, and possibly a representation $r_G$ of an internal global symmetry group. The precise determination 
of the spectrum of an interacting CFT is of fundamental importance. Together with the knowledge of the OPE coefficients, it essentially 
amounts to a solution of the CFT. 

It is well-known that in a unitary CFT$_d$, dimensions of primary operators satisfy certain inequalities known as unitarity bounds 
\cite{Mack:1975je, Minwalla:1997ka}. For a spin $s$ operator $J_{\mu_1\mu_2\cdots \mu_s}$ in the symmetric traceless representation 
of $SO(d)$, the unitarity bound is
\be 
\Delta_s \ge d-2+s\,,\qquad s\ge 1\,.
\label{uBound-s}
\ee
For a scalar operator ${\cal O}$, it reads
\be 
\Delta_0 \ge \frac{d}{2}-1
\ee 
and one may derive similar bounds for more general representations of $SO(d)$. When these inequalities are saturated, the corresponding 
operator satisfies some differential equation, and it belongs to a short representation of the conformal algebra. In the case of a 
scalar operator, the shortening condition is simply the wave equation $\partial^2{\cal O}=0$, i.e. ${\cal O}$ is a free scalar 
field. For a spin $s$ operator ($s\ge 1$), saturation of the bound (\ref{uBound-s}) implies that it is a conserved current 
\be 
\partial^{\mu} J_{\mu \mu_2 \cdots \mu_s}=0\,.
\label{conserv}
\ee 
The cases $s=1$ (an exactly conserved current with $\Delta=d-1$) and $s=2$ (the conserved stress tensor with $\Delta=d$) are familiar in any CFT. 
Conserved currents of higher spins $s>2$ are explicitly realized in free field theories. For example, in a free scalar CFT they take the schematic form 
\be
\label{currentdef}
J_{\mu_1\cdots \mu_s}=\sum^{s}_{k=0} c_{sk} \partial_{\{ \mu_1}\cdots \partial_{\mu_k} \phi  \partial_{\mu_{k+1}}\cdots \partial_{\mu_s \}} \phi\,,
\ee
where brackets denote traceless symmetrization and the coefficients may be determined by the conservation equation, as we 
review in Section 2.1. It is evident that these operators 
have exact dimension $\Delta_s = d-2+s$ in the free theory. As usual, conserved currents correspond to symmetries of the theory. The presence of exactly 
conserved currents of all spins implies that the CFT has an infinite dimensional higher spin symmetry which includes the conformal symmetry as a subalgebra. 
Higher spin symmetries turn out to be very constraining. One may prove that if a CFT possesses a spin 4 conserved current, then an infinite 
tower of conserved higher spin operators is present, and all correlation functions of local operators coincide with those of a free CFT 
\cite{Maldacena:2011jn}.

In an interacting CFT, the higher spin operators are not exactly conserved and acquire an anomalous dimension
\be 
\Delta_s = d-2+s+\gamma_s\,.
\ee 
An interesting class of models are those for which the higher spin symmetries are slightly broken. By this we mean that there is an expansion 
parameter $g$, playing the role of a coupling constant, such that for small $g$ the anomalous dimensions $\gamma_s(g)$ are small, 
and in the $g\rightarrow 0$ limit one recovers exact conservation of the currents. Explicit examples are weakly coupled fixed points 
of the Wilson-Fisher type \cite{Wilson:1971dc}, where $g$ corresponds to a power of $\epsilon$, or certain large $N$ CFT's, 
where $g$ is related to a power of $1/N$. At the operator level, this implies 
that the non-conservation equation for the spin $s$ operator takes the form
\be 
\partial \cdot J_s = g K_{s-1}
\label{non-cons}
\ee  
where $K_{s-1}$ is an operator of spin $s-1$, and we factored out $g$ to highlight the fact that at $g=0$ the current is conserved. 
The slightly 
broken higher spin symmetries (\ref{non-cons}) can still be used to put non-trivial constraints on the correlation functions 
\cite{Maldacena:2012sf}. The equation (\ref{non-cons}) also gives an efficient way to determine the anomalous dimensions $\gamma_s$ to leading 
order in the small parameter $g$ \cite{Anselmi:1998ms, Belitsky:2007jp}. As we review in Section 2.2, using (\ref{non-cons}) and conformal symmetry 
one readily finds that $\gamma_s(g)\propto g^2$, where the proportionality constant is 
simply obtained by computing the two point functions $\langle K_{s-1} K_{s-1}\rangle$ at $g=0$. This method is similar in spirit to 
the one recently advocated in \cite{Rychkov:2015naa}, where the leading anomalous dimension of $\phi$ at the Wilson-Fisher fixed point in $d=4-\eps$ 
was reproduced by using conformal symmetry, without explicit input from perturbation theory. 

In this paper, we apply this method to the explicit example of interacting scalar field theories with $O(N)$ symmetry in various dimensions. These 
include the familiar Wilson-Fisher fixed point of the $\phi^4$ theory in $d=4-\eps$, the large $N$ expansion of the critical $O(N)$ model in arbitrary 
dimension $d$, the perturbative IR fixed points of the cubic $O(N)$ models in $d=6-\eps$ \cite{Fei:2014yja}, and the UV fixed point of the non-linear sigma model 
in $d=2+\eps$. In all these examples, we determine the explicit structure of the non-conservation equation (\ref{non-cons}) and use it to find the leading 
order anomalous dimensions of higher spin operators in the singlet, symmetric traceless and antisymmetric representations of $O(N)$. Many of our 
findings were obtained before by different methods \cite{Wilson:1973jj,Lang:1992zw}, but the results in the cubic models in $d=6-\eps$ and in the nonlinear sigma model in 
$d=2+\eps$ are new as far as we know. In all examples, we pay particular attention to the large spin behavior of the anomalous dimensions, finding 
precise agreement with general expectations \cite{Parisi:1973xn, Callan:1973pu,Alday:2007mf, Komargodski:2012ek,Fitzpatrick:2012yx}. Combining 
information from the $d=4-\eps$ and $d=2+\eps$ expansions, as well as some input from the large spin limit, in Section 7 we also obtain some estimates 
for the dimension of the singlet higher spin operators in the $d=3$ $O(N)$ models for a few low values of spin $s$ and $N$.

In addition to their intrinsic interest and their relevance in statistical mechanics, the $O(N)$ models we study in this note also play an important 
role in the context of the AdS/CFT correspondence. According to a well understood entry of the AdS/CFT dictionary, 
exactly conserved currents of spin $s$ in CFT$_d$ are dual to massless spin $s$ gauge fields in AdS$_{d+1}$. Interacting higher spin gauge 
theories in AdS$_{d+1}$ were explicitly constructed by Vasiliev \cite{Vasiliev:1990en, Vasiliev:1992av, Vasiliev:1999ba, Vasiliev:2003ev}, and 
a class of them were naturally conjectured \cite{Klebanov:2002ja} to be dual to the singlet sector of the free $O(N)$ vector model. The 
exactly conserved currents $J_s$ with $s=2,4,6,\ldots$ are dual to the corresponding massless gauge fields in the Vasiliev theory, and the 
scalar operator $J_0 = \phi^i\phi^i$ to a bulk scalar field with $m^2=-2(d-2)/\ell^2_{AdS}$. As further 
conjectured in \cite{Klebanov:2002ja}, one may extend this duality to the interacting case, obtained by adding to the free theory the ``double-trace" 
interaction $\lambda (\phi^i\phi^i)^2$. For $d<4$ there is a flow to an interacting IR fixed point\footnote{For $4<d<6$ there 
is a flow to a (presumably metastable) perturbatively unitary, UV fixed point \cite{Fei:2014yja}.} which is conjectured to be dual to the same 
Vasiliev theory but with alternate boundary conditions on the bulk scalar field \cite{Klebanov:1999tb}. A distinguishing feature 
of large $N$ interacting vector models is that the descendant operator $K_{s-1}$ appearing in the non-conservation equation is 
a ``double-trace" operator, schematically
\be 
\partial \cdot J_s = \frac{1}{\sqrt{N}}\sum J J\,.
\label{non-cons-vec}
\ee 
This implies that the anomalous dimensions are $\gamma_s \sim O(1/N)$, which corresponds to a quantum breaking of the higher spin gauge symmetry in the bulk: 
the higher spin gauge fields acquire masses through loop corrections\footnote{In the higher spin/CFT dualities, the bulk Newton's constant $G_N$ 
scales as $1/N$. The mass of a spin $s$ field in AdS$_{d+1}$ is related to the dual conformal dimension by 
$(\Delta_s+s-2)(\Delta_s+2-d-s)=m^2_s\ell^2_{AdS}$, 
which implies $m^2_s\ell^2_{AdS} \sim 1/N$ for $\gamma_s \sim 1/N$. To leading order at large $N$, $m^2_s\ell^2_{AdS} \approx (2s+d-4)\gamma_s$.} 
when the bulk scalar is quantized with the alternate boundary conditions.  
In representation
theory language, the equation (\ref{non-cons}) means that the short representation of the conformal algebra with $(\Delta=d-2+s,s)$ combines 
with the representation $(\Delta=d-1+s,s-1)$ to form a long multiplet with $(\Delta > d-2+s,s)$. In the bulk, this phenomenon corresponds to 
a higher spin version of the Higgs mechanism \cite{Girardello:2002pp}: the gauge field swallows a spin $s-1$ Higgs field to yield 
a massive spin $s$ field. The fact that the operator on the right-hand side of (\ref{non-cons-vec}) is double-trace implies that the Higgs field 
is a composite two-particle state, and the breaking is subleading at large $N$. This is different from theories of Yang-Mills type, where 
$K_{s-1}$ in (\ref{non-cons}) is a single trace operator, and the anomalous dimensions are non-zero already at planar level. In the bulk, 
this would correspond to a tree-level Higgs mechanism. 

Let us finally mention that, although in this paper we focus on the scalar $O(N)$ models, the methods we use can be applied to a variety of 
interesting 
theories, such as for instance the critical Gross-Neveu models in $2<d<4$, conformal QED$_{d}$ with $N_f$ fermions, the $CP^{N}$ model, 
and 3d bosonic and fermionic vector models coupled to Chern-Simons gauge fields \cite{Giombi:2011kc, Aharony:2011jz, Maldacena:2012sf}. The latter 
models have approximate higher spin symmetry at large $N$ for all values of the `t~Hooft coupling $\lambda$, and one should be able to use (\ref{non-cons}) 
to find the  anomalous dimensions to order $1/N$, $\gamma_s = f_s(\lambda)/N+\ldots$, for all values of $\lambda$ and spin.\footnote{Some results for 
a few low values of spins were given in \cite{Giombi:2011kc,Maldacena:2012sf}.} We leave this to future work. 

{\it Note added in proof:} After completion of our calculations, while writing up this note, 
we became aware of \cite{Skvortsov:2015pea}, which has overlap with some of our results. 

\section{General method}
\subsection{The higher spin currents in free field theory}
In this section we will setup the definitions and notations which will then be applied to the particular models. 
We will also describe the derivation of the master formula which allows to calculate the lowest-order value of the anomalous dimensions without doing any loop calculations.

We first introduce some useful technology for the manipulation of symmetric tensors. For a given a rank $s$ tensor $J_{\mu_1\mu_2\cdots \mu_s}$
in the symmetric traceless representation, we may introduce an auxiliary ``polarization vector" $z^{\mu}$, which can be taken to be null ($z^2=0$), 
and construct the index-free projected tensor
\be
\hat{J}_s \equiv  J_{\mu_1\cdots \mu_s} z^{\mu_1}\cdots z^{\mu_s}\,,\qquad z^2=0\,.
\ee
It is evident that the multiplication by $z^{\mu}$ selects only the symmetric traceless part of $J_{\mu_1\cdots \mu_s}$. One may always go 
back to the full tensor by ``stripping off" the null vectors and subtracting traces. In practice, this can be done efficiently 
with 
with the help of the following differential operator in $z$-space \cite{Dobrev:1975ru, Belitsky:2007jp, Costa:2011mg}
\be
D^{\mu}_z\equiv \left(\frac{d}{2}-1\right)\partial_{z_\mu} +z^\nu \partial_{z_\nu} \partial_{z_\mu} - \frac{1}{2} z^\mu \partial_{z_\nu} \partial_{z_\nu}.
\ee
Acting once with this operator removes a $z^{\mu}$, thus freeing one index of the tensor, while taking into account the constraint $z^2=0$. The 
unprojected $J_{\mu_1\cdots \mu_s}$ can thus be recovered via
\be 
J_{\mu_1 \mu_2 \cdots \mu_s} \propto D^z_{\mu_1}D^z_{\mu_2}\cdots D^{z}_{\mu_s} \hat{J}_s \,.
\ee 
The symmetrization and tracelessness of the operator obtained this way is ensured by the properties
\be
[D^{\mu}_z,D^{\nu}_z]=0,\qquad  D^{\mu}_z D_{\mu}^z=0.
\ee
Similarly, the conservation equation (\ref{conserv}) of the spin $s$ operator may be written compactly in this notation as
\be
\label{conservproj}
\partial_{\mu} D^{\mu}_z\hat{J}_s=0.
\ee

Let us now construct the explicit conserved higher spin currents in the free CFT of $N$ real massless scalar fields. They satisfy 
the free wave equation
\be 
\partial^2 \phi^i = 0, \qquad i=1,\ldots,N
\ee 
and there is a $O(N)$ global symmetry under which $\phi^i$ transforms in the fundamental representation. This free CFT admits an infinite 
tower of exactly conserved higher spin operators (\ref{currentdef}), which are bilinears in the scalars with a total of $s$ derivatives 
acting on the fields. Projecting indices with the null vector $z^{\mu}$, these operators can be written as 
\be 
\hat{J}^{ij}= \sum^{s}_{k=0} c_{sk} \deh^{s-k} \phi^i \deh^{k} \phi^j
\label{Jhat}
\ee 
where we have introduced the projected derivative $\hat \partial = \partial_{\mu} z^{\mu}$, and $c_{sk}$ are coefficients that will be fixed shortly. 
Of course, one can separate this operator into irreducible representations of 
$O(N)$, as discussed in more detail below. It is convenient to rewrite (\ref{Jhat}) in the following form 
\begin{equation}
\begin{gathered}
\label{definition}
\hat{J}^{ij}_s=f_{s}(\deh_1,\deh_2)\phi^{i}(x_1)\phi^j(x_2)\Big{|}_{x_1,x_2\rightarrow x} \\
f_s(u,v)=\sum^{s}_{k=0} c_{sk} u^{s-k} v^{k}\,,\qquad  u=\deh_1,v=\deh_2.
\end{gathered}
\end{equation}
where we have encoded the coefficients $c_{sk}$ into the function $f_s(u,v)$. Now the conservation equation (\ref{conservproj}) 
may be turned into a differential equation for the function $f_s$, which, upon using the free equation of motion $\partial^2\phi=0$, reduces to
\begin{equation}
\label{scalarfeq}
\big((d/2-1)(\partial_u+\partial_v) + u \partial_{u}^2  + v \partial_{v}^2\big) f_{s}=0.
\end{equation}
The following ansatz for $f_s$ is convenient
\begin{equation}
f_s=(u+v)^s \phi_s\Big(\frac{u-v}{u+v}\Big),
\end{equation}
and results in the ordinary differential equation
 \begin{equation}
\Big((1-t^2)\frac{d^2}{dt^2}-(d-2)t\frac{d}{dt} + s(s+d-3)\Big)\phi_s(t)=0\,.
\end{equation}
The solution to this equation is given by the order $s$ Gegenbauer polynomials, $\phi_s(t) = C^{d/2-3/2}_s(t)$,  which are even (odd) for even (odd) $s$.
Hence, up to the overall normalization, one gets the following expressions for the conserved higher spin currents
\begin{equation}
\label{formula}
\hat{J}^{ij}_s=(\deh_1+\deh_2)^{s}C^{d/2-3/2}_{s}\Big(\frac{\deh_1-\deh_2}{\deh_1+\deh_2}\Big)\phi^{i}(x_1)\phi^j(x_2)\Big{|}_{x_{1},x_{2}\rightarrow x}.
\end{equation} 
One may also write
\begin{eqnarray}
\label{Jsum}
&&(u+v)^{s}C^{d/2-3/2}_{s}\left(\frac{u-v}{u+v}\right)=\\
&&\qquad =\frac{\sqrt{\pi } \Gamma \left(\frac{d}{2}+s-1\right) \Gamma (d+s-3)}{2^{d-4}\Gamma \left(\frac{d-3}{2}\right)}\sum_{k=0}^s 
\frac{(-1)^k u^{s-k} v^k}{k!(s-k)!\,\Gamma\left(k+\frac{d}{2}-1\right)\Gamma\left(s-k+\frac{d}{2}-1\right)}\nonumber
\end{eqnarray}
from which one can read-off the coefficients $c_{sk}$ in (\ref{Jhat}) if desired. The overall normalization is arbitrary at this level. Note that 
one feature of the form (\ref{formula}) is that it vanishes at $d=3$, see the factor in front of the sum in (\ref{Jsum}). This vanishing is not 
meaningful, one could always remove it by normalizing the currents differently. For the explicit calculations below, we find it more convenient 
to use the form (\ref{formula}) in terms of Gegenbauer polynomials. 

The higher spin operators may be decomposed into symmetric traceless, antisymmetric and singlet of $O(N)$
\be 
J_s^{ij} = J_s^{(ij)}+J_s^{[ij]}+J_s
\label{allcurrents}
\ee 
where $J_s\equiv J_s^{ii}$ denotes the singlet current. It is evident by symmetry that the singlet and symmetric traceless representations only exist for even spin, and the antisymmetric one 
for odd spins. For $s=1$, the antisymmetric operator $J_1^{[ij]}$ is nothing but the familiar conserved current corresponding 
to the $O(N)$ global symmetry. The presence of the conserved currents of all spins implies that the free CFT 
has an infinite dimensional exact higher spin symmetry. The generators 
can be constructed in a canonical way as follows. First, by contracting a spin $s$ current with a spin $s-1$ conformal Killing tensor 
$\zeta^{\mu_1 \ldots \mu_{s-1}}$,\footnote{A conformal Killing tensor is a symmetric tensor satisfying 
$\partial_{(\mu_1}\zeta_{\mu_2\cdots \mu_{s})}=\frac{s-1}{d+2s-4} g_{(\mu_1\mu_2} \partial^{\nu} \zeta_{\mu_3\cdots \mu_s)\nu}$.}
we may obtain an ordinary current $J_{\mu,s}^{\zeta} = J_{\mu \mu_2\cdots \mu_s}\zeta^{\mu_2 \ldots \mu_{s}}$, which is conserved 
as it is easily checked. From this, one can get a conserved charge $Q_s^{\zeta}$ in the usual way. For instance, for $s=2$, the 
singlet current $J_2$ is proportional to the traceless stress tensor of the CFT,
and contracting this with the linearly independent conformal Killing vectors one gets the $(d+2)(d+1)/2$ generators of the conformal algebra. 
In the interacting theory, all of the currents (\ref{allcurrents}), except for $J_1^{[ij]}$ and $J_2$, will be broken.  In particular, 
while the free CFT has $N(N+1)/2$ conserved ``stress tensors", only one of them remains conserved when interactions are switched on. 

For what follows, it will be useful to work out the normalization of the two point function of the currents (\ref{formula}) in arbitrary dimensions $d$. 
Since the currents are bilinear in $\phi$ there will be two propagators, which are differentiated by the hatted derivatives at both points. 
The calculation is drastically simplified by using the Schwinger parametrization of the propagator
\begin{equation}
\label{schwinger}
\langle \phi^i(x) \phi^{j}(0) \rangle= \frac{\Gamma(d/2-1)}{4\pi^{d/2}}\frac{\delta^{ij}}{(x^2)^{d/2-1}}=\delta^{ij}\int^{\infty}_{0}\frac{d\alpha}{4\pi^{d/2}} \alpha^{d/2-2}e^{-\alpha x^2}\,.
\end{equation}
Owing to the fact that $\deh \hat{x} = 0$, since $z^2=0$, all hatted derivatives are replaced by $-2\alpha \hat{x}$ if acting at point $x$ and $+2\alpha \hat{x}$ at point $0$, so that instead of spin sums we have integrals of Gegenbauer polynomials over the parameters $\alpha_1$ and $\alpha_2$ for the first and second propagator respectively. Separating the $O(N)$ indices, we may write
\be
\langle \hat{J}^{ij}_s(x) \hat{J}^{kl}_s(0) \rangle =(\delta^{ik}\delta^{jl}  +(-1)^s \delta^{il} \delta^{jk}) 
\frac{ {\cal N}_s (\hat{x})^{2s}}{(x^2)^{d+2s-2}}
\ee
The $(-1)^s$ comes from the property of $C^{d/2-3/2}_s(-x) = (-1)^s C^{d/2-3/2}_s(x)$. The spacetime and $z^{\mu}$ dependence is 
of course as required by conformal symmetry for a spin $s$ conserved operator. The normalization factor ${\cal N}_s$ 
is given by the following expression
\be
\label{currentnorm}
\begin{gathered}
\frac{(-1)^s 2^{2s}}{(4\pi^{d/2})^2}\int^{\infty}_{0} \int^{\infty}_{0} d\alpha_1d\alpha_2 \alpha_1^{d/2-2} \alpha_2^{d/2-2} e^{-\alpha_1-\alpha_2}(\alpha_1+\alpha_2)^{2s} C^{d/2-3/2}_{s}\Big(\frac{\alpha_1-\alpha_2}{\alpha_1+\alpha_2}\Big) C^{d/2-3/2}_{s}\Big(\frac{\alpha_1-\alpha_2}{\alpha_1+\alpha_2}\Big)\\
=\frac{(-1)^s 2^{2s}}{(4\pi^{d/2})^2}\int^{\infty}_{0} dp \frac{p^{d-4+2s+1}}{2^{d-3}} e^{-p} \int^{1}_{-1} dq (1-q^2)^{d/2-2} (C^{d/2-3/2}_{s} (q))^2\\
\end{gathered}
\ee
from which we get
\be 
\label{Ns}
{\cal N}_s= \frac{(-1)^s 2^{2s}}{(4\pi^{d/2})^2} \frac{\pi \Gamma(d+2s-3)\Gamma(d+s-3)}{2^{2d-8}s!(\Gamma(d/2-3/2))^2} 
\ee 
The norms corresponding to the irreducible representations of $O(N)$ are then
\be 
\label{JJ-singlet}
\langle \hat{J}_s(x) \hat{J}_s(0) \rangle = N(1+(-1)^s)  
\frac{{\cal N}_s(\hat{x})^{2s}}{(x^2)^{d+2s-2}} 
\ee
for the singlet,
\be
\langle \hat{J}^{(ij)}_s(x) \hat{J}^{(kl)}_s(0) \rangle =  \frac{(1+(-1)^s)}{2} (\delta^{ik}\delta^{jl}  + \delta^{il} \delta^{jk}-\frac{2}{N}\delta^{ij}\delta^{kl} )
\frac{{\cal N}_s(\hat{x})^{2s}}{(x^2)^{d+2s-2}}  
\ee
for the symmetric traceless, and
\be
\langle \hat{J}^{[ij]}_s(x) \hat{J}^{[kl]}_s(0) \rangle =  \frac{(1-(-1)^s)}{2} (\delta^{ik}\delta^{jl}  -\delta^{il} \delta^{jk})  
\frac{ {\cal N}_s(\hat{x})^{2s}}{(x^2)^{d+2s-2}} 
\ee
for the antisymmetric. 

\subsection{Anomalous dimensions of the weakly broken currents}
Let us consider a CFT with a parameter $g$ playing the role of a coupling constant, such that in the $g=0$ limit 
there are exactly conserved currents $J_s$. 
When a non-zero coupling $g$ is turned on, the currents will be no longer conserved for general $s$ and acquire anomalous dimensions
\be
\Delta_s = d-2 + s+ \gamma_s(g)\,.
\ee
The non-conservation of the currents means that a non-zero operator of spin $s-1$ must appear on the right hand side of (\ref{conserv}), 
or equivalently (\ref{conservproj})
\be
\label{descgen}
\partial_{\mu} D^{\mu}_z \hat{J}_s=g\hat{K}_{s-1}\,,
\ee
where we have pulled out an explicit factor of $g$ in front of the descendant to stress that the right hand side vanishes when $g=0$. Here 
$g$ is assumed to be a small expansion parameter, and may be either a power of $\epsilon$ in the Wilson-Fisher type models, or a power of $1/N$ 
in the large $N$ approach. We now proceed by noting that in a CFT the form of the two-point function of 
the spin $s$ operators is fixed by conformal symmetry to be
\be
\label{conftwopoint}
\langle \hat{J}_s (x_1) \hat{J}_{s'}(x_2)\rangle = \delta_{ss'} C(g) \frac{\hat{I}^{s}}{(x^2_{12})^{\Delta_{s}}}
\ee
where 
\be
\hat{I} = I_{\mu\nu}z_1^\mu z_2^\nu\,,\qquad  I_{\mu\nu} = \eta_{\mu\nu} - 2\frac{x_{12}^\mu x_{12}^\nu}{x^2_{12}} \,.
\ee
Acting on this two-point function with $\partial_\mu D^{\mu}_z$ on both operators (with different projection vectors $z_1$ and $z_2$), one gets, 
using the form of the non-conservation equation (\ref{descgen})
\be
\partial_{1\mu}D^{\mu}_{z_1} \partial_{2\mu}D^{\mu}_{z_2} \langle \hat{J}_s (x_1) \hat{J}_{s}(x_2)\rangle = g^2 \langle \hat{K}_{s-1} (x_1) \hat{K}_{s-1}(x_2)\rangle.
\ee
On the other hand, differentiating the right hand side of (\ref{conftwopoint}), setting $z_1=z_2$ at the end, 
and dividing by the two-point function of $J's$, one finds the relation \cite{Anselmi:1998ms, Belitsky:2007jp}
\begin{multline}
\label{mainform}
g^2 \hat{x}^2 \frac{\langle \hat{K}_{s-1} (x_1) \hat{K}_{s-1}(x_2)\rangle}{\langle \hat{J}_s (x_1) \hat{J}_{s}(x_2)\rangle}= -\gamma_s(g^2) s (s+d/2-2) \big[ (s+d/2-1)(s+d-3) \\+ \gamma_s(g^2) (s^2+sd/2-2s+d/2-1)\big].
\end{multline}
The right-hand side being proportional to $\gamma_s$ is not a coincidence and follows from the conservation of the higher-spin current at zero coupling (\ref{conservproj}). From a CFT standpoint, (\ref{mainform}) is an exact relation. In practice, when doing perturbation theory in $g$, one computes the 
correlators on the left hand side in powers of the coupling. It is then evident that (\ref{mainform}) allows to gain an order in perturbation 
theory. To obtain the anomalous dimensions of the broken currents to leading order in $g$, one has simply to evaluate ratio of correlators 
in the free theory, $g=0$. In particular, this only involves finite tree-level correlators, avoiding 
the issues of regularization and renormalization. 

The considerations above are general and apply to any CFT with weakly broken higher spin operators (\ref{descgen}). 
For the explicit examples discussed in the rest of the paper, it will be useful to determine the general form of the descendants $K_{s-1}$ in 
the scalar theories. Applying the divergence operator to the higher spin currents (\ref{formula}), we find in terms of the function $f_s(u,v)$:
\be
\begin{gathered}
\label{descendant}
\partial_{\mu} D^{\mu}_z \hat{J}_s^{ij} = \big[h_s(\deh_1,\deh_2) \partial_1^2 +(-1)^s h_s(\deh_2,\deh_1) \partial_2^2 \big] \phi^i(x_1)\phi^j(x_2)
\Big{|}_{x_1,x_2\rightarrow x},\\
h_s(u,v) \equiv (d/2-1) \partial_u f_s+\frac{u-v}{2} \partial^2_{uu} f_{s}+v \partial^2_{uv} f_s\,.
\end{gathered} 
\ee
Of course, this is zero in the free theory where $\partial^2_1=\partial^2_2=0$. In the interacting theory, (\ref{descendant}) allows to determine 
the form of the descendent once the equation of motion for $\phi$ is known in the specific model of interest.
The function $h_s(u,v)$ can be evaluated more explicitly using the recurrence relations between the Gegenbauer polynomials, and one finds
\be
h_s(u,v) = (u+v)^{s-1} (d-3)\big[(d/2-1) C_{s-1}^{d/2-1/2}\Big(\frac{u-v}{u+v}\Big)-\frac{2(d-1)v}{u+v} C_{s-2}^{d/2+1/2}\Big(\frac{u-v}{u+v}\Big)\big]\,.
\label{hs}
\ee
As discussed above, the vanishing at $d=3$ is superficial and is a consequence of the normalization of the currents (\ref{formula}). 

Note that the methods described in this section can also be used to fix the leading order anomalous dimension of a nearly free field. 
For instance, in the case of a scalar field, the equation of motion takes the form
\be 
\partial^2 \phi = g V
\ee 
where $V$ is some operator of spin zero and bare dimension $d/2+1$. By an analogous calculation as the one described above for the higher spin operators, 
one can show that to leading order $\gamma_{\phi}\propto g^2$, where the proportionality constant is related to the two point function $\langle VV\rangle$ 
at $g=0$. We will use this method in the next Section to reproduce the well-known anomalous dimension of $\phi$ at the Wilson-Fisher fixed point, 
see also \cite{Rychkov:2015naa}. The analogous calculations 
in the large $N$ approach and in the nonlinear sigma model in $d=2+\epsilon$ are given in Section 4 and 6 respectively.

\section{$O(N)$ model in $d=4-\epsilon$}
We now apply the general formulae obtained in the previous section to the case of the critical $O(N)$ $\phi^4$ model in $d=4-\epsilon$ dimensions, 
with action 
\begin{equation}
\label{actionONep}
S=\int d^d x\Big(  \frac{1}{2} \partial_{\mu} \phi^{i}  \partial^{\mu} \phi^{i}+ \frac{\lambda}{4}(\phi^i \phi^i)^2\Big).
\end{equation}
The one-loop beta function is well known and reads
\be
\beta(\lambda) = -\eps \lambda + \frac{(N+8)\lambda^2}{8\pi^2}
\ee
and thus there is a IR critical point at $\lambda_{*}=\frac{8\pi^2}{N+8}\epsilon+O(\epsilon^2)$. Before moving on to the higher spin operators, 
let us show how to reproduce the leading order anomalous dimension of $\phi$ using the classical equations of motion, following the methods reviewed above, 
see also \cite{Rychkov:2015naa}.  
In the free theory, when $\lambda=0$, the elementary field $\phi^i$ has canonical dimension $\Delta_0 = d/2 -1$, thus saturating the unitarity bound 
and obeying $\partial^2 \phi^i =0$. The tree-level two-point function of $\phi^i$ is given by
 \be
 \label{tree}
\langle \phi^i (x_1) \phi^j(x_2) \rangle = \frac{\Gamma(d/2-1)}{4\pi^{d/2}}\frac{\delta^{ij}}{(x_{12}^2)^{d/2-1}}.
\ee
When we turn on the interaction, the equation of motion is modified to
\be
\label{nonconserv}
 \partial^2 \phi^i = \lambda \phi^i \phi^j \phi^j,
\ee
and the two-point function receives corrections. At the conformal point, 
the exact two-point function is constrained by conformal symmetry to be 
 \be
 \label{sc2fn}
\langle \phi^i (x_1) \phi^j(x_2) \rangle = \delta^{ij}\frac{C(\lambda)}{(x_{12}^2)^{d/2-1+\gamma_{\phi}}}.
\ee
Applying the equation of motion twice, i.e. taking the $\partial^2_1 \partial^2_2$ on both sides and taking the ratio one gets:
\be
\label{trick}
\lambda_*^{2} (x^2_{12})^2 \frac{\langle \phi^i \phi^k \phi^k (x_1)  \phi^j \phi^l \phi^l(x_2) \rangle}{\langle \phi^i (x_1) \phi^j(x_2) \rangle} = 4\gamma_{\phi} (\gamma_{\phi}+1)(d-2+2\gamma_{\phi})(d+2\gamma_{\phi}).
\ee
The fact that the right-hand side is proportional to $\gamma_{\phi}$ is expected and is due to the shortening condition at zero coupling, $\partial^2 \phi^i =0$. To get the leading order in $\eps$ for $\gamma_{\phi}$ we notice that in the left hand side $\lambda_*^{2}\sim \eps^2$, so in the two-point function ratio we can just plug $d=4$ propagators
\be
\langle \phi^i (x_1) \phi^j(x_2) \rangle = \frac{1}{4\pi^2}\frac{\delta^{ij}}{x_{12}^2}.
\ee
In the right-hand side we get
\be
\label{rhs}
4\gamma_{\phi}(\gamma_{\phi}+1)(2-\eps+\gamma_{\phi})(4-\eps+2\gamma_{\phi}) = 32\gamma_{\phi}+O(\eps^3)
\ee
since it is evident that $\gamma_{\phi}\sim\eps^2$. Now, the two-point function in the numerator of (\ref{trick}),
evaluated at tree level, yields
\be
\langle \phi^i \phi^k \phi^k (x_1)  \phi^j \phi^l \phi^l(x_2) \rangle_0 = \frac{1}{(4\pi^2)^3}(2N+4)\delta^{ij} \frac{1}{(x^2_{12})^3}
\ee
Finally, taking the ratio by the free propagator and equating the right-hand side (\ref{rhs}), we recover the well-known result
\be
\gamma_{\phi}=\frac{\lambda_*^{2}}{(4\pi^2)^2}\frac{N+2}{16} =\frac{N+2}{4(N+8)^2}\epsilon^2.
\label{gamphi4d}
\ee
One can contrast this to the usual calculation, which is technically quite different. There, the leading order correction to the two-point function of the $\phi^i$ field is given by extracting the logarithmic divergence of the standard two-loop diagram in Fig. 1. 
\begin{figure}
                \centering
                \includegraphics[width=3cm]{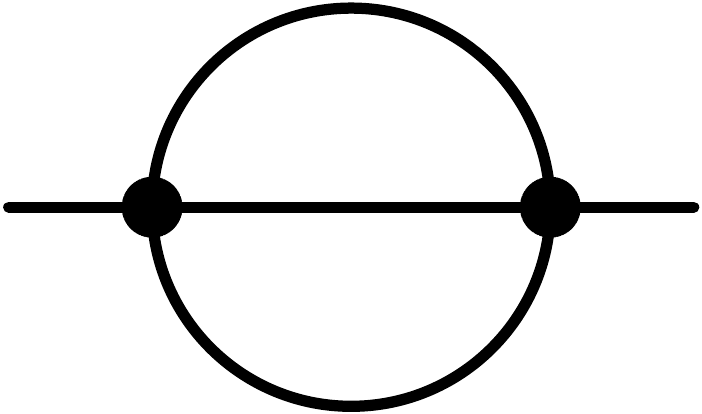}
               \label{phi-2loop}
               \caption{The 2-loop diagram yielding the leading order anomalous dimension of the $\phi^i$ field in the $O(N)$ model in the standard 
approach.}
\end{figure}

We may now proceed to studying the higher spin currents using the same method. We use the definition of the currents (\ref{formula}) and the descendant (\ref{descendant}). To lowest order the $\eps$ dependence is fixed by the the critical coupling $\lambda_{*}$, so we can use $d=4$ everywhere. The currents are then:
\be
\hat{J}^{ij}_s=(\deh_1+\deh_2)^{s}C^{1/2}_{s}\Big(\frac{\deh_1-\deh_2}{\deh_1+\deh_2}\Big)\phi^{i}(x_1)\phi^j(x_2)\Big{|}_{x_1,x_2\rightarrow x},
\ee
and the descendant:
\begin{equation}
\label{desceps}
\begin{gathered}
\hat{K}^{ij}_{s-1}(x)= \Big(h_s(\deh_1+\deh_3+\deh_4,\deh_2)+(-1)^s h_s(\deh_2+\deh_3+\deh_4,\deh_1)\Big) \phi^i(x_1) \phi^j(x_2) \phi^k(x_3) \phi^k(x_4)
\Big{|}_{x_{1,2,3,4}\rightarrow x}\,, \\
h_s (u,v) = (u+v)^{s-1} \big[C_{s-1}^{3/2}\Big(\frac{u-v}{u+v}\Big)-\frac{6v}{u+v} C_{s-2}^{5/2}\Big(\frac{u-v}{u+v}\Big)\big] \\
\end{gathered}
\end{equation}
Note that this form is redundant in the sense that we could combine $\deh_3+\deh_4$ into $\deh_3$ acting on $\phi^i \phi^i (x_3)$, but it makes all the symmetries of the diagrams we will need to calculate explicit. A few examples might be useful. For instance, for $s=1$ the only non-zero current is the antisymmetric one, for which the descendant vanishes as it should since it's the current in the adjoint of $O(N)$. For $s=2$ we have non-zero currents for the symmetric traceless and the singlet representations, and the descendant is
\be
\hat{K}_1^{ij}=2(\deh_3+\deh_4-\deh_1-\deh_2)\phi^i(x_1) \phi^j(x_2) \phi^k(x_3) \phi^k(x_4)\Big{|}_{x_{1,2,3,4}\rightarrow x} = -2\deh (\phi^i \phi^j)\phi^k \phi^k + 2 \phi^i \phi^j\deh (\phi^k \phi^k)
\ee 
This vanishes for the singlet as it is the conserved stress-energy tensor. It does not vanish for the symmetric traceless representation, as 
the corresponding operator acquires an anomalous dimension in the interacting theory. As another example, the 
spin 3 descendant of the spin 4 singlet current is
\begin{eqnarray}
&&\hat{K}_3 = 20 ( \phi^i \phi^i \deh^3 (\phi^k \phi^k)  - 6 \deh  (\phi^i \phi^i ) \deh^2 (\phi^k \phi^k) +33( \phi^i \deh^2 \phi^i  - 30 \deh \phi^i \deh \phi^i )\deh  (\phi^k \phi^k)\cr 
&&\qquad \qquad  - 3\deh (3\phi^i \deh^2 \phi^i -4\deh \phi^i  \deh \phi^i )\phi^k \phi^k
\end{eqnarray}


The master formula (\ref{mainform}) at the leading order yields the following for $\gamma_s$: 
\begin{equation}
\label{dim}
\gamma_s = - \frac{\lambda_*^{2}}{s^2(s+1)^2} \frac{\hat{x}^2 \langle  \hat{K}_{s-1}  (x) \hat{K}_{s-1} (0)\rangle}{\langle  \hat{J}_s  (x) \hat{J}_s (0)\rangle}\,.
\end{equation}
For the two-point function of the currents one has in $d=4$, according to (\ref{currentnorm}):
\be
\langle \hat{J}_s \hat{J}_s \rangle = 2N  \frac{1}{(4\pi^{2})^2}  \frac{(2\hat{x})^{2s}}{(x^2)^{2s+2}} \frac{\pi \Gamma(2s+1)\Gamma(s+1)}{s!(\Gamma(1/2))^2} 
= 2N \frac{(2s)!}{(4\pi^2)^2} \frac{(2\hat{x})^{2s}}{(x^2)^{2s+2}} 
\label{JJ4d-singlet}
\ee
for the singlet and similarly 
\begin{eqnarray}
\langle \hat{J}^{(ij)}_s \hat{J}^{(kl)}_s \rangle =  (\delta^{ik}\delta^{jl}  + \delta^{ik} \delta^{jl}-\frac{2}{N}\delta^{ij}\delta^{kl} )   
\frac{(2s)!}{(4\pi^2)^2} \frac{(2\hat{x})^{2s}}{(x^2)^{2s+2}} 
 \\
\langle \hat{J}^{[ij]}_s \hat{J}^{[kl]}_s \rangle =   (\delta^{ik}\delta^{jl}  -\delta^{ik} \delta^{jl})
\frac{(2s)!}{(4\pi^2)^2} \frac{-(2\hat{x})^{2s}}{(x^2)^{2s+2}} 
\label{JJ4d-symmAsymm}
\end{eqnarray}
where we used the fact that the singlet and symmetric traceless representations exist for even spins only, and the antisymmetric one for odd spins. 

To obtain the anomalous dimensions via eq.~(\ref{dim}), we have to compute the two-point function of the descendant at tree level. Each descendant 
(\ref{desceps}) consists of a differential operator acting on four $\phi$ fields. We simply have to compute the free field Wick contractions between the fields 
(contractions of fields on the same descendant are of course excluded)
\begin{equation}
\langle \phi^i(x_1)\phi^j(x_2)\phi^m(x_3)\phi^m(x_4)\,,\phi^k(y_1)\phi^l(y_2)\phi^n(y_3)\phi^n(y_4)\rangle_0
\end{equation}
and then act with the differential operator in (\ref{desceps}) on the resulting product of free propagators, 
setting $x_{1,2,3,4}\rightarrow x$ and $y_{1,2,3,4}\rightarrow 0$ at the end. It is straightforward to do this for any given spin: the problem 
is purely algebraic and there are no integrals to compute. However, 
to obtain a general result as a function of spin, it is convenient to use the Schwinger representation (\ref{schwinger}) of the propagator and 
carry out the resulting integrals of products of Gegenbauer polynomials. Some technical details of this are collected in Appendix \ref{app-tech}. 
The final result takes the following form. 
For even spins, based on symmetry we get the structure
\begin{equation}
\label{symKij}
\langle \hat K^{ij}_{s-1} \hat K^{kl}_{s-1} \rangle = (A_s N + C_s) \frac{\delta^{ik}\delta^{jl} + \delta^{il}\delta^{jk}}{2} + B_s \delta^{ij} \delta^{kl}\,,
\qquad s=2,4,6,\ldots
\end{equation}
and similarly for odd spins
\begin{equation}
\label{asymKij}
\langle \hat K^{ij}_{s-1} \hat K^{kl}_{s-1} \rangle = (A'_s N + C'_s) \frac{\delta^{ik}\delta^{jl} - \delta^{il}\delta^{jk}}{2}\,,
\qquad s=1,3,5,\ldots 
\end{equation}
The $A_s$ (and $A_s'$) terms come from contracting the first pair of $\phi$ fields at different points with each other and the second pair as well (hence the $O(N)$ indices form a closed loop); the $B_s$ term is from contracting the pairs across (pair one with pair two and vice-versa); the $C_s$ (and $C_s'$) term is from contracting one $\phi$ from the first pair with one from the second one (so that the $O(N)$ indices thread the diagram without loops). The final result for the coefficients $A_s, B_s, C_s, A'_s, C'_s$ 
is:
\be
\begin{aligned}
A_s &= -(s-1)s(s+1)(s+2)(2s)! \frac{(2\hat{x})^{2s-2}}{(x^2)^{2s+2}(4\pi^2)^4}\\
B_s &= 4s(s+1)(2s)! \frac{(2\hat{x})^{2s-2}}{(x^2)^{2s+2}(4\pi^2)^4}\\
C_s &= -2(s-2)s(s+1)(s+3)(2s)! \frac{(2\hat{x})^{2s-2}}{(x^2)^{2s+2}(4\pi^2)^4}\\
A'_s&=-A_s \,,\qquad C'_s=2A_s
\end{aligned}
\ee

From (\ref{symKij}) and (\ref{asymKij}), we can readily extract the singlet, symmetric traceless and antisymmetric parts. They are 
\begin{equation}
\begin{aligned}
&\langle \hat K_{s-1} \hat K_{s-1}\rangle = (A_s+B_s)N^2+C_s N\\
&\langle \hat K_{s-1}^{(ij)} \hat K_{s-1}^{(kl)}\rangle = \frac{A_s N+C_s}{2} \left(\delta^{ik}\delta^{jl}  + \delta^{ik} \delta^{jl}-\frac{2}{N}\delta^{ij}\delta^{kl}\right)\\
&\langle \hat K_{s-1}^{[ij]} \hat K_{s-1}^{[kl]}\rangle = -\frac{A_s}{2}(N+2)\left(\delta^{ik}\delta^{jl}  -\delta^{ik} \delta^{jl}\right)\,.
\label{KKirreducible}
\end{aligned}
\end{equation}
It is now straightforward to extract the one-loop anomalous dimensions, using the general formula (\ref{dim}) and the normalization 
of the currents (\ref{JJ4d-singlet}), (\ref{JJ4d-symmAsymm}). For the singlet operators, we get\footnote{It is amusing that the spin 
dependent factor in brackets in (\ref{gamma4e}) is the same as the central charge of the unitary minimal models $M(s,s+1)$, which have 
$c=0,1/2,7/10,\ldots$ for $s=2,3,4, \ldots$. Similarly, the result (\ref{gammaAsym}) for the antisymmetric representation is proportional 
to the central charge $c=\frac{3}{2}\left(1-8/(p(p+2))\right)$ of the ${\cal N}=1$ supersymmetric minimal models for $p=2s$. One may wonder if there is a deeper significance to these apparent coincidences. We thank Igor Klebanov for bringing this to our attention.}
\be
\label{gamma4e}
\gamma_s = \frac{\eps^2(N+2)}{2(N+8)^2}\Big(1-\frac{6}{s(s+1)}\Big)
\ee
This vanishes for $s=2$ as it should, corresponding to the conservation of the stress-energy tensor. 
For the symmetric traceless operators, we get
\begin{equation}
\label{gamma4sym}
\gamma_{s(ij)}= \frac{\epsilon^2 (N+2)}{2(N+8)^2}\Big(1-\frac{2(N+6)}{(N+2)s(s+1)}\Big)
\end{equation}
and for the antisymmetric ones 
\begin{equation}
\label{gammaAsym}
\gamma_{s[ij]}= \frac{\epsilon^2 (N+2)}{2(N+8)^2}\Big(1-\frac{2}{s(s+1)}\Big)\,.
\end{equation}
The latter vanishes for $s=1$, corresponding to the exact conservation of the current in the adjoint of $O(N)$. All these results 
are in agreement with \cite{Wilson:1973jj}.

It is worth mentioning the analysis of these results done in \cite{Komargodski:2012ek,Fitzpatrick:2012yx}, where the asymptotic $s\rightarrow\infty$ behavior for the twist, $\tau_s=\Delta_s-s$ and consequently the anomalous dimension of higher-spin currents is established. Namely one should get for the twist of the higher spin operator of the form $\phi\deh^s\phi$  
\begin{equation}
\label{zhibform}
 \tau_s= 2\tau_{\phi}-\frac{c_{\tau_{min}}}{s^{\tau_{min}}}+\ldots,
\end{equation}
 where $\tau_{min}$ is the minimal twist among the operators present in the $\phi\phi$ OPE. First of all, we see that the  limiting value is equal to $\frac{\eps^2(N+2)}{2(N+8)^2}=2\gamma_{\phi}$ as follows from expanding (\ref{zhibform}) to order $\eps$. Second, the leading correction behaves as $\frac{1}{s^2}$, which is a manifestation of having a tower of operators with twist 2, which at this order in $\eps$ are the higher-spin currents and the $\phi^i\phi^i$ operator (see \cite{Komargodski:2012ek} for more details). In principle, the coefficient $c_{\tau_{min}}$ is determined by the certain three-point functions of these operators with $\phi$. We will go into more detail about this in the next two sections.

\section{The large $N$ critical $O(N)$ model}
To develop the $1/N$ expansion of the $\phi^4$ theory, one may introduce a Hubbard-Stratonovich auxiliary field $\sigma$, so that 
the action (\ref{actionONep}) may be rewritten as
\be
\label{actionlarge}
S=\int d^d x\Big(  \frac{1}{2} \partial_{\mu} \phi^{i}  \partial^{\mu} \phi^{i}+ \frac{1}{2} \sigma \phi^i \phi^i -\frac{\sigma^2}{4\lambda}\Big).
\ee
In the IR limit for $d<4$, the last term becomes unimportant and can be dropped.\footnote{For $d>4$, the last term can be dropped in the UV limit, 
corresponding to a non-trivial UV fixed point.} To develop perturbation theory, it is convenient to rescale $\sigma$ so that the action becomes
\be
\label{actionlargeN}
S=\int d^d x\Big(  \frac{1}{2} \partial_{\mu} \phi_{i}  \partial^{\mu} \phi^{i}+ \frac{1}{2\sqrt{N}} \sigma \phi^i \phi^i\Big).
\ee
The $\sigma$ field then acquires an effective non-local propagator upon integrating out $\phi^i$\footnote{The quadratic term in the resulting 
$\sigma$ effective action is just proportional to the two-point function $\langle \phi^i\phi^i(x)\phi^j\phi^j(y)\rangle_0$ in the free theory. 
The $\sigma$ propagator is obtained by Fourier transforming to momentum space, inverting, and transforming back to $x$-space.} 
\be
\langle \sigma(x_1) \sigma(x_2) \rangle = \frac{C_{\sigma\sigma}}{(x^2_{12})^2}\,,\qquad  C_{\sigma\sigma}=\frac{2^{d+2}\Gamma(\frac{d}{2}-\frac{1}{2})\sin(\pi d/2)}{\pi^{\frac{3}{2}}\Gamma(\frac{d}{2}-2)}
\ee
so that $\sigma$, which replaces the scalar operator $\phi^i\phi^i$, is a primary operator of dimension $2+O(1/N)$ at the interacting fixed point. 
Systematic perturbation theory can be developed using this effective propagator, 
the canonical propagator  (\ref{tree}) for $\phi^i$ and the $\sigma \phi^i\phi^i$ vertex, with $1/\sqrt{N}$ playing the role of the coupling constant.

The equation of motion for $\phi$ is
\be
\label{largeNeq}
\partial^2 \phi^i = \frac{1}{\sqrt{N}}\sigma \phi^i
\ee
and the equation of motion for $\sigma$ is formally $\phi^i\phi^i=0$ after we drop the last term of (\ref{actionlargeN}) in the IR limit. The role of this equation is to ``subtract" from the theory 
the operator $\phi^i\phi^i$, which is replaced by $\sigma$. This fact will play an important role in our calculation below. 

Before turning to the higher spin currents let us calculate as a warmup the anomalous dimension $\gamma_{\phi}$ of the $\phi$ field, 
without computing Feynman diagrams. Using the equations of motion (\ref{largeNeq}) and acting on the $\phi$ two-point function with 
$\partial^2_1\partial^2_2$, we get
\be
\frac{1}{N} (x^2_{12})^2 \frac{\langle \phi^i \sigma (x_1)  \phi^j \sigma(x_2) \rangle}{\langle \phi^i (x_1) \phi^j(x_2) \rangle} = 2\gamma_{\phi} (\gamma_{\phi}+1)(d-2+2\gamma_{\phi})(d+2\gamma_{\phi}).
\ee
From this, the leading order value of $\gamma_{\phi}$ immediately follows
\be
\gamma_{\phi} = \frac{C_{\sigma\sigma}}{4Nd(d-2)}=\frac{2\sin(\pi d/2 ) \Gamma(d-2)}{N \pi \Gamma(\frac{d}{2}-2) \Gamma(\frac{d}{2}+1)} 
\ee
which is a well-known result. 
It is quite remarkable how simple the calculation is, provided we know $C_{\sigma\sigma}$. It is also helpful that one completely avoids (to the lowest
order) the issues of regularization and renormalization, which are actually somewhat thorny in the $1/N$ expansion.

Let us now turn to the higher spin currents. These have the same form (\ref{formula}), and using (\ref{descendant}) and the equations of motion 
(\ref{largeNeq}), one ends up with
\begin{equation}
\partial_{\mu}D^{\mu}_z \hat J_s^{ij} = \frac{1}{\sqrt{N}} \hat K_{s-1}^{ij}
\end{equation}
where
\begin{equation}
\begin{gathered}
\label{desc1n}
\hat{K}_{s-1}^{ij}=  \Big(h_s(\deh_1+\deh_3,\deh_2)+(-1)^s h_s(\deh_2+\deh_3,\deh_1)\Big) \phi^i(x_1) \phi^j(x_2) \sigma(x_3)\Big{|}_{x_{1,2,3}\rightarrow x}\,,
\end{gathered}
\end{equation}
and the function $h_s(u,v)$ is given in eq.~(\ref{hs}).\footnote{Note that the vanishing of $h_s$ at $d=3$ is of course not meaningful, 
and just follows from the normalization of the currents that we have chosen. Such factors of $(d-3)$ cancel out in the 
ratio $\langle K_{s-1}K_{s-1}\rangle /\langle J_s J_s\rangle$.} 
It is possible, and convenient for what follows, to decompose the descendant in products of the conformal primaries 
$\hat{J}_s$, $\sigma$ and their derivatives. We find
\begin{equation}
\begin{gathered}
\hat{K}_{s-1}^{ij} = \sum_{s'=0}^{s-2} \sum_{k=0}^{s-s'-1} C_{s'k} \deh^{s-s'-k-1} \hat{J}_{s'}^{ij} \deh^k \sigma 
\label{K-naive}
\end{gathered}
\end{equation}
where the coefficients $C_{s'k}$ are given explicitly by
\be
C_{s'k}=  
\begin{cases}(s-s') (2s'+d-3)  \binom{s-s'-1}{k} \binom{-s-s'+k-d+3}{k+1}\,,\quad s-s' ~{\rm even}\\
0\,,\qquad\qquad \qquad \qquad \qquad ~~~~~~~~~~~~~~~~~~~~~~~~\,\, s-s'~{\rm odd} 
\end{cases}
\ee
An important point is that so far we have only used the $\phi$ equation of motion (\ref{largeNeq}), 
and not the equation for $\sigma$ whose role is to formally project out $J_0 = \phi^i\phi^i$ from the theory. This implies that in fact the 
form of the descendant (\ref{K-naive}) only applies as written to the non-singlet currents. For the singlets, one obtains the correct 
descendant by the prescription that the term $s'=0$ should be dropped from the sum. As an example, for $s=2$ we get from (\ref{K-naive})
\be
\hat{K}_1^{ij} = (d-1)(d-3) \big((d-2) \hat{J}_0^{ij} \deh \sigma - 2 (\deh \hat{J}_0^{ij}) \sigma \big)
\ee
For the symmetric traceless $K_{1}^{(ij)}$, on the right hand side we have $\hat{J}_0^{(ij)}=\phi^{(i} \phi^{j)}$ 
and the descendant is non-vanishing. However, for the singlet we would have $J_0=\phi^i\phi^i$, which should be thrown away. This 
leads to $\hat{K}_1=0$, as it should be according to the conservation of the stress-energy tensor. In a similar way, 
the spin 3 descendant of the spin 4 singlet current is
\be
\hat{K}_3 = (d+3)(d+1) \big((d+2) \hat{J}_2 \deh \sigma  - 2  (\deh \hat{J}_2) \sigma \big)
\ee
At $d=3$, we see that $\hat{K}_3 \propto \hat{J}_2 \deh \sigma -2/5 (\deh \hat{J}_2) \sigma$, in agreement with \cite{Maldacena:2012sf}. 

We can now compute the anomalous dimensions using (\ref{mainform}). Let us first discuss the case of the non-singlet currents, where we can 
use directly the form (\ref{desc1n}), equivalent to (\ref{K-naive}). The descendant two-point function can be computed similarly to the previous section
and one ends up with
\begin{equation}
\label{zhibsym}
\gamma_{s(ij)}= \gamma_{s[ij]}=  2\gamma_{\phi} \frac{(s-1)(d+s-2)}{(d/2+s-2)(d/2+s-1)},
\end{equation}
where $\gamma_{\phi}$ is the anomalous dimension of $\phi$ field. This is the correct result, in agreement \cite{Lang:1992zw}. Let us now turn to the case of the singlet currents, which is slightly more involved due to the $J_0$ 
projection discussed above. The correct singlet descendant is given by (here as usual we denote 
by $J_{s'}=J_{s'}^{ii}$ the singlet operators)
\begin{equation}
\begin{aligned}
&\hat K_{s-1} = \sum_{s'=2}^{s-2} \sum_{k=0}^{s-s'-1} C_{s'k} \deh^{s-s'-k-1} \hat{J}_{s'} \deh^k \sigma  = 
\hat K_{s-1}^{\rm naive}-\hat K^{0}_{s-1}\\
&\hat K^{0}_{s-1}=\sum_{k=0}^{s-1} C_{0k} \deh^{s-k-1} \hat{J}_{0}\deh^k \sigma
\end{aligned}
\end{equation}
where $\hat K_{s-1}^{\rm naive}$ coincides with (\ref{desc1n}), with $O(N)$ indices traced. Its two-point function leads to a contribution 
equal to (\ref{zhibsym}) to the anomalous dimension. To subtract the contribution of the term $\hat K^{0}_{s-1}$ proportional to $J_0$, we note 
that
\begin{equation}
\langle \hat K_{s-1}^{\rm naive} \hat K_{s-1}^{\rm naive}\rangle = \langle \hat K_{s-1} \hat K_{s-1}\rangle +\langle \hat K_{s-1}^{0} \hat K_{s-1}^{0}\rangle 
\end{equation}
since $\hat K_{s-1}$ and $\hat K_{s-1}^{0}$ are orthogonal, due to orthogonality of the spin $s$ primaries ($\langle J_s J_{s'}\rangle \sim 
\delta_{ss'}$). Then, the correct singlet anomalous dimension 
is obtained by simply subtracting from (\ref{zhibsym}) the contribution of the two-point function $\langle \hat K_{s-1}^{(0)} \hat K_{s-1}^{(0)}\rangle$.  
By this procedure, we get the final result
\begin{equation}
\label{zhib}
\gamma_s=2\gamma_{\phi}  \frac{1}{(d/2+s-2)(d/2+s-1)}\Big[(s-1)(d+s-2) - \frac{\Gamma(d+1)
\Gamma(s+1)}{2(d-1)\Gamma(d+s-3)}\Big]\,,
\end{equation}
in agreement with \cite{Lang:1992zw}. One may also check that setting $d=4-\eps$, (\ref{zhib}) and (\ref{zhibsym}) precisely match the 
the results (\ref{gamma4e}), (\ref{gamma4sym}), (\ref{gammaAsym}) expanded to order $1/N$.  

It is again interesting to mention the $s\rightarrow \infty$ behavior \cite{Komargodski:2012ek, Fitzpatrick:2012yx}. The expansion at large $s$ yields
\be
\begin{gathered}
\label{LargeNInfSpin}
\gamma_s = 2\gamma_{\phi} - 2\gamma_{\phi}\frac{\Gamma(d+1)}{2(d-1)}\frac{1}{s^{d-2}} - 2\gamma_{\phi}\frac{d(d-2)}{4}\frac{1}{s^2} +...,\\
\gamma_{s(ij)}= 2\gamma_{\phi}  - 2\gamma_{\phi}\frac{d(d-2)}{4}\frac{1}{s^2}+...
\end{gathered}
\ee
We see a tower of higher spin currents ($1/s^{d-2}$) in the singlet channel, as well as the $\sigma$ ($1/s^2$) in both the singlet and the traceless parts. The higher-spin contribution vanishes for the symmetric traceless. The $1/s^2$ coming from $\sigma$ is universal and can be calculated. The coefficient $c_{\tau_{min}}$ in (\ref{zhibform}) is given by the formula (3.18) of \cite{Komargodski:2012ek} which reads in terms of two- and three-point function coefficients
\begin{equation}
\label{zhibcoef}
c_{\tau_{min}}=\frac{\Gamma(\tau_{min}+2s_{min})}{2^{s_{min}-1}\Gamma(\frac{\tau_{min}+2s_{min}}{2})^2}\frac{\Gamma(\Delta)^2}{\Gamma(\Delta-\frac{\tau_{min}}{2})^2} \frac{C_{OOO_{\tau_{min}}}^2}{C_{OO}^2 C_{O_{\tau_{min}}O_{\tau_{min}}}}
\end{equation}
In this theory, $O=\phi$, $O_{\tau_{min}} = \sigma$. Here $\Delta = d/2-1$, $\tau_{min}=2$, $C_{O_{\tau_{min}}O_{\tau_{min}}}$ is the three-point function coefficient of $\phi \phi \sigma$, $C_{\phi\phi}=\frac{\Gamma(d/2-1)}{4\pi^{d/2}}$, $C_{\sigma\sigma}$ is defined above. Plugging all the factors, the coefficient of $1/s^2$ is exactly reproduced.

\section{Cubic models in $d=6-\epsilon$}
let us consider the following model with $N+1$ scalars and $O(N)$ invariant cubic interactions
\be
S= \int d^d x\Big(  \frac{1}{2} \partial_{\mu} \phi^{i}  \partial^{\mu} \phi^{i}+ \frac{1}{2} (\partial_{\mu} \sigma)^2  + \frac{g_1}{2} \sigma \phi^i \phi^i + \frac{g_2}{6} \sigma^3\Big). 
\ee
As argued in \cite{Fei:2014yja}, in $d=6-\epsilon$ this model posseses IR stable, perturbatively unitary fixed points which provide a ``UV completion" 
of the large $N$ UV fixed points of the $O(N)$ model in $d>4$. This proposal has passed various non-trivial checks 
\cite{Fei:2014yja, Fei:2014xta, Gracey:2015tta}. These perturbative fixed points exist for $N>1038(1+O(\eps))$, and are expected to be unitary to all 
orders in $\epsilon$ and $1/N$ expansions. However, non-perturbative effects presumably render the vacuum metastable via instanton effects. 
In this section we just perform perturbative calculations, and in particular we obtain further non-trivial agreement with the large $N$ expansion in $d>4$.
\footnote{Calculations of anomalous dimensions of higher spin operators in a similar cubic model with an adjoint scalar and $\tr\phi^3$ interaction 
were carried out in \cite{Braun:2013tva,Manashov:2015fha}.}

In this model, a novel feature in the calculation of the anomalous dimension of the higher spin operators is that the free theory contains 
two independent towers of conserved, $O(N)$ singlet, higher spin currents:
\begin{equation}
\begin{gathered}
\hat{J}_{s,\phi} = \frac{1}{\sqrt{N}} (\deh_1+\deh_2)^{s}C^{3/2}_{s}\Big(\frac{\deh_1-\deh_2}{\deh_1+\deh_2}\Big)\phi^i (x_1)\phi^i (x_2)
\Big{|}_{x_1,x_2\rightarrow x} \\
\hat{J}_{s,\sigma} = (\deh_1+\deh_2)^{s}C^{3/2}_{s}\Big(\frac{\deh_1-\deh_2}{\deh_1+\deh_2}\Big)\sigma (x_1) \sigma (x_2)
\Big{|}_{x_1,x_2\rightarrow x} 
\end{gathered}
\end{equation}
where we have normalized $J_{s,\phi}$ so that both currents have $\langle J_s J_s\rangle \sim O(1)$. Once interactions are turned on, we 
expect these operators to mix non-trivially, and one should determine the appropriate eigenstates of the dilatation operator. 

The equations of motion are 
\begin{equation}
\begin{gathered}
\partial^2 \phi^i = g_1 \sigma \phi^i \\
\partial^2 \sigma = \frac{1}{2} (g_1 \phi^i \phi^i + g_2 \sigma^2)\,.
\label{eom6}
\end{gathered}
\end{equation}
It is evident that the equations of motion will induce the mixing between the currents, 
since $\langle \partial \cdot J_{s,\phi} \partial \cdot  J_{s,\sigma}\rangle \neq 0$ due to the $g_1$-dependent interactions 
in (\ref{eom6}). The descendant operators (in this case we find it more convenient to include the coupling constants 
into the definition of the $K_{s-1}$'s)
\begin{equation}
\partial_{\mu}D^{\mu}_z \hat{J}_{s,\phi}=\hat{K}_{s-1,\phi}  \,,\qquad \partial_{\mu}D^{\mu}_z \hat{J}_{s,\sigma}=\hat{K}_{s-1,\sigma} 
\end{equation}
can be computed in a straightforward way by following similar steps as in the previous sections. Explicitly, they are given by 
\begin{eqnarray}
&&\hat K_{s-1,\phi} =  \frac{1}{\sqrt{N}}\left(h_s(\deh_1+\deh_3,\deh_2)+1 \leftrightarrow 2\right)g_1\phi^i(x_1)\phi^i(x_2) \sigma (x_3)
\Big{|}_{x_{1,2,3}\rightarrow x}\\
&&\hat K_{s-1,\sigma} = \left(h_s(\deh_1+\deh_3,\deh_2)+1 \leftrightarrow 2\right)\left(
\frac{g_1}{2}\phi^i(x_1)\sigma(x_2)\phi^i(x_3)+\frac{g_2}{2}\sigma(x_1) \sigma(x_2)\sigma(x_3)\right)\Big{|}_{x_{1,2,3}\rightarrow x}\cr
&&h_s (u,v) = 6(u+v)^{s-1} \big[C_{s-1}^{5/2}\Big(\frac{u-v}{u+v}\Big)-\frac{5v}{u+v} C_{s-2}^{7/2}\Big(\frac{u-v}{u+v}\Big)\big]
\nonumber
\end{eqnarray}
We can then use the general relation (\ref{mainform}), 
suitably generalized to the present case with non-trivial mixing, to obtain the following anomalous dimension mixing matrix
\begin{equation}
\begin{bmatrix}
   \frac{g^2_1}{48\cdot4\pi^3}(1-\frac{6}{(s+1)(s+2)}) & -\frac{g^2_1}{4\pi^3} \frac{1}{8(s+1)(s+2)}  \sqrt{N}  \\
   -\frac{g^2_1}{4\pi^3} \frac{1}{8(s+1)(s+2)}  \sqrt{N} &  \frac{g^2_1}{4\pi^3}\frac{1}{2\cdot 48} N+\frac{g^2_2}{2\cdot 48\cdot 4\pi^3}(1-\frac{12}{(s+1)(s+2)})
\end{bmatrix}
\label{mixingM}
\end{equation}
where the non-diagonal terms comes from the non-zero two-point function $\langle \hat{K}_{s-1,\phi} \hat{K}_{s-1,\sigma}\rangle$. From this mixing matrix, 
one can compute the two eigenvalues to leading order in $\epsilon$ and finite $N$, using the expression for the fixed point couplings given in \cite{Fei:2014yja}. 
The resulting finite $N$ expressions are easy to get, but rather lengthy. At large $N$, using the expressions for the fixed point couplings \cite{Fei:2014yja}
\begin{align}
g_{1}^* &= \sqrt{\frac{6\epsilon (4\pi)^3}{N}}\left(1+\frac{22}{N}+\frac{726}{N^2}-\frac{326180}{N^3}+\ldots\right)\label{g1star}  \\
g_{2}^* &= 6 \sqrt{\frac{6\epsilon (4\pi)^3}{N}}\left(1+\frac{162}{N}+\frac{68766}{N^2}+\frac{41224420}{N^3}+\ldots\right)\label{g2star}
\end{align}
one finds that the eigenvalues are given by
\begin{equation}
\begin{gathered}
\label{gammacub}
\gamma_1 = \frac{2\epsilon}{N}\frac{(s-2)(s+5)(s^2+3s+8)}{(s+1)^2(s+2)^2}+O(1/N^2)\\
\gamma_2 = \epsilon+\frac{16\epsilon}{N}\frac{5s^4+30s^3+38s^2-21s-25}{(s+1)^2(s+2)^2}+O(1/N^2)
\end{gathered}
\end{equation}
The higher order corrections can be obtained to any desired order, but for simplicity we have listed here only the leading order in $1/N$. We 
see that the $\gamma_1$ eigenvalue vanishes at $s=2$, and one can check that this is true for any $N$. This eigenvalue then corresponds to the 
tower of ``single-trace" higher spin currents which include the stress-energy tensor. Indeed, one can explicitly verify that $\gamma_1$ matches 
the $1/N$ expansion result (\ref{zhib}) expanded in $d=6-\epsilon$. The dimension corresponding to the second eigenvalue is 
\begin{equation}
\Delta_2 = d-2+s+\gamma_2 = 4+s+\frac{16\epsilon}{N}\frac{5s^4+30s^3+38s^2-21s-25}{(s+1)^2(s+2)^2}+O(1/N^2)
\label{delta2}
\end{equation}
which suggests that this should match the ``double-trace" operator $ \sigma \partial^s \sigma \sim \phi^2 \partial^s \phi^2$ in the large $N$ 
approach. Indeed, one can match (\ref{delta2}) with the result given in \cite{Lang:1992zw} for the dimension of such composite operators of spin $s$. 
For $s=0$, this is the scalar operator $\sigma^2$ of the large $N$ model, which has dimension $\Delta = 4-100\eps/N+\ldots$ near $d=6$ and corresponds 
to a particular mixture of the mass operators $\phi^i\phi^i$ and $\sigma^2$ in the cubic model \cite{Fei:2014yja}. 

One can also study the spin $s$ operators in the symmetric traceless and antisymmetric representations of $O(N)$, where no mixing occurs (since $\hat{J}_{s,\sigma}$ is a singlet of $O(N)$). Following similar steps to the previous sections, we obtain the result
\begin{equation}
\gamma_{s(ij)}= \gamma_{s[ij]}=  \frac{(g_1^*)^2}{192\pi^3}\frac{(s-1)(s+4)}{(s+1)(s+2)}
=\frac{2\epsilon}{N}(1+\frac{44}{N}+\ldots)\frac{(s-1)(s+4)}{(s+1)(s+2)}
\label{non-singlet}
\end{equation}
The order $1/N$ is seen to exactly match the large $N$ result (\ref{zhibsym}). Furthermore, we checked that the $1/N^2$ term also matches with the 
result obtained in \cite{Derkachov:1997ch} using large $N$ methods for arbitrary $d$.\footnote{As far as we know, the $1/N^2$ term in the large $N$
expansion of the anomalous dimensions of the {\it singlet} higher spin operators has not been obtained in the literature.}

Let us now study the large spin limit of these results. For the eigenvalues of the singlet mixing matrix, the large spin expansion 
can be written in a simple form, valid for finite $N$, in terms of the fixed point couplings 
\begin{equation}
\begin{gathered}
\label{gammacub}
\gamma_1 = \frac{(g_1^*)^2}{192\pi^3} - \frac{(g_1^*)^2}{32\pi^3} \frac{1}{s^2}+\ldots , = 2\gamma_{\phi} - \frac{(g_1^*)^2}{32\pi^3} \frac{1}{s^2}+\ldots\\
\gamma_2 = \frac{(g_1^*)^2 N + (g_2^*)^2 }{384\pi^3} - \frac{(g_2^*)^2}{32\pi^3} \frac{1}{s^2}+\ldots 
=2\gamma_{\sigma} - \frac{(g_2^*)^2}{32\pi^3} \frac{1}{s^2}+\ldots 
\end{gathered}
\end{equation}
where we have used the known expression for the one-loop anomalous dimensions of $\phi$ and $\sigma$ in the cubic model \cite{Fei:2014yja}. The leading 
terms are precisely consistent with the expected large spin limit. The subleading $1/s^2$ contributions are clearly coming from the exchange 
of the $\sigma$ field, which has $\Delta_\sigma=\tau=2$. We can check explicitly the prediction of the formula (\ref{zhibcoef}) 
for the coefficients of the $1/s^2$ terms. For the $\gamma_1$ eigenvalue,  we should take that $O=\phi,O_{\tau_{min}} = \sigma$. The two-point function coefficients are $C_{\phi \phi}$=$C_{\sigma \sigma}=\frac{1}{4\pi^3}$. The three point function coefficient is given at the lowest order by a diagram with one $g_1 \phi^i\phi^i$ vertex in the middle. The diagram is given by the integral: 
\begin{equation}
\frac{g_1}{(4\pi^3)^3}\int \frac{d^6 x_0}{x_{10}^4 x_{20}^4 x_{30}^4} =  \frac{g_1}{(4\pi^3)^3}\frac{\pi^3}{x_{12}^2 x_{23}^2 x_{31}^2}
\end{equation}
Combining the factors we get $c_{\tau_{min}}= 2\frac{(g_1^*)^2}{(4^3\pi^6)^2}(4\pi^3)^3=\frac{(g_1^*)^2}{32\pi^3}$. Overall, we then get 
\begin{equation}
\tau_{s,\phi} = d - 2 + \frac{(g_1^*)^2}{192\pi^3} - \frac{(g_1^*)^2}{32\pi^3}\frac{1}{s^2}+\ldots = 2\tau_\phi - \frac{(g_1^*)^2}{32\pi^3}\frac{1}{s^2}+\ldots
\end{equation}
since $\tau_\phi=\Delta_\phi=d/2 - 1 + \frac{(g_1^*)^2}{384\pi^3}$ in the leading order. The same applies for the eigenvalue $\gamma_2$, 
corresponding to $\tau_{s,\sigma}$, where we get: 
\begin{equation}
\tau_{s,\sigma} = d - 2 + \frac{(g_1^*)^2 N + (g_2^*)^2 }{384\pi^3} - \frac{(g_2^*)^2}{32\pi^3} \frac{1}{s^2}+\ldots 
= 2 \tau_\sigma - \frac{(g_2^*)^2}{32\pi^3} \frac{1}{s^2}+\ldots
\end{equation}
where $\tau_\sigma=\Delta_{\sigma} = d/2 - 1 + \frac{(g_1^*)^2 N + (g_2^*)^2 }{768\pi^3}$ is the leading order dimension (and thus twist) of $\sigma$. The coefficient of $1/s^2$ is reproduced the three-point function $\langle \sigma \sigma\sigma\rangle$.

The cubic model in $d=6-\epsilon$ also admits non-unitary fixed points which are of interest in statistical mechanics. The simplest case 
is the $N=0$ model, which just consists of a single scalar field $\sigma$ with cubic interaction $g_2/6 \sigma^3$. 
This model has a non-unitary fixed point at
\begin{equation}
(g_2^*)^2 =-\frac{128\pi^3}{3}\eps+O(\eps^2)
\end{equation}
As pointed out by Fisher \cite{Fisher:1978pf}, this theory is related to the Lee-Yang edge singularity of the Ising model. For $d=2$ ($\eps=4$), the 
fixed point corresponds to the non-unitary minimal model $M(2,5)$. Using the result (\ref{mixingM}) for $g_1=0, N=0$, we can deduce the dimension 
of the higher spin operators $\sim \sigma \partial^s\sigma$ in the Fisher model to be
\begin{equation}
\gamma_s = 2\gamma_{\sigma}
\left(1-\frac{12}{(s+1) (s+2)}\right)= -\frac{\eps}{9}  \left(1-\frac{12}{(s+1) (s+2)}\right)+O(\eps^2)\,.
\end{equation}
where we have used $\gamma_{\sigma} = \frac{(g_2^*)^2}{768\pi^3}$ at one loop.

Another interesting non-unitary model is obtained at the formal value $N=-2$. In this case the model is equivalent to a theory 
of a complex anticommuting scalar $\theta$ and a commuting scalar $\sigma$ \cite{Fei:2015kta}
\be
S=\int d^d x \bigg (
\partial_\mu \theta \partial^\mu \bar \theta + \frac{1}{2}\left(\partial_{\mu}\sigma\right)^2 +g_1 \sigma \theta \bar \theta + \frac{1}{6}g_2 \sigma^3
\bigg )
\ .\ee
with $Sp(2)$ global symmetry. 
The IR stable fixed point occurs at \cite{Fei:2015kta}
\begin{equation}
g_2^*=2 g_1^*\,,\qquad g_1^* = i\sqrt{\frac{(4\pi)^3\epsilon}{5}}\left(1+O(\epsilon)\right)\ ,
\label{N2-fixed}
\end{equation}
where the first equality holds to all orders in perturbation theory. For such a relation between couplings, one can verify that the model has an 
enhanced ``supersymmetry" $OSp(1|2)$ which implies that the dimension of $\theta$ and $\sigma$ are equal. It turns out that this $OSp(1|2)$ invariant 
fixed point is equivalent to the $q\rightarrow 0$ limit of the $q$-state Potts model \cite{Caracciolo:2004hz}. The dimension of the $Sp(2)$ invariant 
higher spin currents at the fixed point can be obtained from (\ref{mixingM}) setting $N=-2$ and $g_2=2 g_1$. This yields the two eigenvalues
\begin{equation}
\begin{aligned}
&\gamma_1 =\frac{(g_1^*)^2}{192\pi^3} \frac{(s-2) (s+5)}{(s+1) (s+2)} = -\frac{\eps (s-2) (s+5)}{15 (s+1) (s+2)}  \\
&\gamma_2 = \frac{(g_1^*)^2}{192 \pi ^3 }\frac{(s (s+3)-16)}{(s+1) (s+2)}=-\frac{\eps  (s (s+3)-16)}{15 (s+1) (s+2)}
\end{aligned}
\end{equation}
We see that the first eigenvalue corresponds to the tower which includes the stress tensor of the theory, since it vanishes at $s=2$ 
(this eigenvalue corresponds to an $OSp(1|2)$ singlet). In the large spin limit one gets
\begin{equation}
\begin{aligned}
&\gamma_1=-\frac{\eps}{15}+\frac{4 \eps}{5 s^2}+\ldots\\ 
&\gamma_2=-\frac{\eps}{15}+\frac{6 \eps}{5 s^2}+\ldots\,.
\end{aligned}
\end{equation}
The equality of the leading terms is a consequence of $\Delta_{\theta}=\Delta_{\phi}$, as follows from the $OSp(1|2)$ symmetry. One may also obtain 
the dimension of the non-singlet currents, which are the same as in (\ref{non-singlet}), with $g_1$ given in (\ref{N2-fixed}).

\section{Nonlinear sigma model}
It is well established that the critical behavior of the $O(N)$ $\phi^4$ model can be related to the critical nonlinear sigma model, 
see e.g. \cite{Moshe:2003xn} for a review. One of the ways to understand this relation is via the $1/N$ expansion, which provides an explicit 
``interpolation" between the UV fixed points of the sigma model in $d=2+\eps$ and the IR fixed points of the $\phi^4$ model in $d=4-\eps$. 
In this section, we calculate the anomalous dimensions of the higher-spin currents at the critical point of the sigma model in $d=2+\epsilon$, at 
finite $N$. As far as we know, this result has not been obtained elsewhere.

We start with the action with an auxiliary field inserted to resolve the sphere constraint on the $\phi^{i}$ field, $\phi^i \phi^i = 1/g^2$
\be
\label{SigmaAction}
S= \int d^d x\Big(  \frac{1}{2} \partial_{\mu} \phi^{i}  \partial^{\mu} \phi^{i} + \sigma (\phi^i \phi^i -\frac{1}{g^2}) \Big).
\ee
To develop perturbation theory, one may resolve the constraint by introducing a set of $N-1$ independent fields. A convenient parametrization 
is
\begin{equation}
\begin{gathered}
\label{constraint}
\phi^{a} = \varphi^{a},\qquad  a=1,\ldots, N-1;\\
\phi^N = \frac{1}{g} \sqrt{1-g^2 \varphi^a \varphi^a} = \frac{1}{g} - \frac{g}{2} \varphi^a \varphi^a + O(g^3).
\end{gathered}
\end{equation}
In terms of the $\varphi^a$ fields, the action is
\begin{equation}
\label{SigmaCartesian}
S = \int d^d x\Big(  \frac{1}{2} \partial_{\mu} \varphi^{a}  \partial^{\mu} \varphi^{a}
+\frac{g^2}{2}\frac{(\varphi^a\partial_{\mu}\varphi^a)^2}{1-g^2\varphi^a\varphi^a}\Big)
\end{equation}
To leading order in perturbation theory and in $d=2+\epsilon$, the coupling constant has the beta function
\be 
\beta = \frac{\epsilon}{2}g-(N-2)\frac{g^3}{4\pi}
\ee 
and there is a UV fixed point at \cite{Brezin:1975sq,Bardeen:1976zh}
\be 
g^2_{*}=\frac{2\pi\eps}{N-2}
\ee 
The factor of $N-2$ is due to the fact that the $O(2)$ model is conformal and has a trivial beta function in $d=2$. Consequently, the 
perturbative UV fixed point in $d=2+\epsilon$ only exists for $N>2$. 

Before moving onward to the higher spin operators, we will calculate the anomalous dimension of the $\varphi$ field (or equivalently $\phi^i$ in the 
action (\ref{SigmaAction})) using the classical equations of motion. To leading order in $g$, they are given by\footnote{These may be also obtained 
starting from the equations of motion coming from (\ref{SigmaAction}), which are $\partial^2 \phi^i = - g^2  \phi^i \partial_\mu \phi^j \partial_\mu \phi^j$, and resolving the constraint by (\ref{constraint}).}
\begin{equation}
\label{sigmaeq}
\partial^2 \varphi^a = - g^2  \varphi^a \partial_\mu \varphi^b \partial_\mu \varphi^b+O(g^4)\,.
\end{equation}
In full analogy with the discussion in $d=4-\epsilon$, we can apply the equations of motion (\ref{sigmaeq}) to the two-point function of $\varphi$ field, 
obtaining
\begin{equation}
\label{gammaSigma}
g^4 x^4\frac{\langle \varphi^a (\partial_\mu \varphi^c)^2 (x)  \varphi^b (\partial_\mu \varphi^d)^2 (0)  \rangle}{\langle \varphi^a(x) \varphi^b(0) \rangle} = 
4 \gamma_{\phi} (\gamma_{\phi}+1) (d-2+2\gamma_{\phi})(d+2\gamma_{\phi})\,.
\end{equation}
To specialize to the expansion in $d-2=\epsilon$ we need to mention several important points: $\gamma$ will be of the order $\epsilon$ and not $\epsilon^2$, 
unlike in $d=4-\epsilon$. This means that the third term in the right-hand side is of the order $\epsilon$ as well. Second, the bare propagator is 
\be
\langle \varphi^a(x) \varphi^b (0) \rangle = \frac{\Gamma(\frac{\epsilon}{2})}{4\pi \pi^{\epsilon/2}(x^2)^{\epsilon}},
\ee
and applying derivatives to it produces powers of $\epsilon$. They will combine with $\Gamma(\epsilon/2)= \frac{2}{\epsilon}+O(1)$. Since $g_{*}^{4} \sim \epsilon^2$ at the critical point, only terms of order $\epsilon^2$ are needed from the two-point function in the numerator, since we have three $\Gamma's$ on top and one on bottom, which amounts to $1/\epsilon^2$. 
Having said that, the relevant term of the two-point function is easy to calculate. The first and the second derivatives of the propagator are:
\begin{eqnarray}
\partial_\mu\frac{1}{(x^2)^{\epsilon/2}} &=& - \epsilon \frac{x_\mu}{(x^2)^{\epsilon/2}},\\
\label{eqprop}
 \partial_\nu\partial_\mu \frac{1}{(x^2)^{\epsilon/2}}&=& \frac{-\epsilon}{(x^2)^{\epsilon/2+1}}\big(\delta_{\mu\nu}-\frac{2x_\mu x_\nu}{x^2}\big)+ \epsilon^2 \frac{x_\mu x_\nu}{(x^2)^{\epsilon/2+2}}.
\end{eqnarray}
It is evident then that the only way to get $O(\epsilon^2)$ is to contract $\varphi^a$ and $\varphi^b$ which would be undifferentiated, and the other $\varphi$'s 
accordingly so that the $\epsilon$ from (\ref{eqprop}) is picked up two times. Overall one gets for the left hand side of (\ref{gammaSigma})
\begin{equation}
g_*^4 (N-1)\frac{2\cdot2\cdot\epsilon^2}{(4\pi)^2}\frac{2^2}{\epsilon^2}=(N-1)\frac{4\epsilon^2}{(N-2)^2}\,.
\end{equation}
The right-hand side of (\ref{gammaSigma}) yields $8\gamma_{\phi}(\epsilon+2\gamma_{\phi})$ to leading order in $\epsilon$. 
Solving the resulting quadratic equation for $\gamma_{\phi}$ gives 
\begin{equation}
\label{sigmaanom}
\gamma_{\phi}= \frac{\epsilon}{2(N-2)},
\end{equation}
which is the well-known result \cite{Brezin:1975sq, Bardeen:1976zh}. 

Let us now move to the higher spin operators, restricting to the case of the $O(N)$ singlets. The form 
of the higher spin currents is most easily written in terms of the constrained fields $\phi^i, i=1,\ldots, N$ appearing in 
the action (\ref{gammaSigma}). In terms of these fields, they take the same form (\ref{formula})
\begin{equation}
\label{sigmacur}
\hat{J}_s=(\deh_1+\deh_2)^{s}C^{-1/2}_{s}\Big(\frac{\deh_1-\deh_2}{\deh_1+\deh_2}\Big)\phi^{i}(x_1)\phi^i(x_2)\Big{|}_{x_1,x_2\rightarrow x}.
\end{equation}
where we have set $d=2$ since we will only perform a leading order calculation. 

It turns out that due to the properties of $C^{-1/2}_{s}(x)$, in the currents (\ref{sigmacur}) all terms have both $\phi^{i}(x_1)$ and $\phi^{i}(x_2)$ differentiated 
at least once, so that after resolving the constraint (\ref{constraint}), we have in terms of $\varphi^a$
\be
\hat{J}_s=(\deh_1+\deh_2)^{s}C^{-1/2}_{s}\Big(\frac{\deh_1-\deh_2}{\deh_1+\deh_2}\Big)\big(\varphi^{a}(x_1)\varphi^{a}(x_2)+ \frac{g^2}{4} \varphi^{a}\varphi^{a}(x_1) \varphi^{b}\varphi^{b}(x_2) \big)\Big{|}_{x_1,x_2\rightarrow x} +O(g^4)
\label{J-cartesian}
\ee
One may check, for instance, that for $s=2$ this yields the correct stress tensor coming from (\ref{SigmaCartesian}). The reason that we 
have to keep the term of order $g^2$ is that, when we compute the descendant by $\partial_{\mu}D^{\mu}_z\hat J_s$, 
both terms in (\ref{J-cartesian}) yield a contribution of order $g^2$ (because the first term is a conserved current at $g=0$, but the second is not). 
Using the general equation (\ref{descendant}), we have (recall that $s$ is even)
\be 
\partial_{\mu}D^{\mu}_z\hat J_s = \left(h_s(\hat\partial_1,\hat\partial_2)\partial_1^2+h_s(\hat\partial_2,\hat\partial_1)\partial_2^2\right)
\left(\varphi^{a}(x_1)\varphi^{a}(x_2)+ \frac{g^2}{4} \varphi^{a}\varphi^{a}(x_1) \varphi^{b}\varphi^{b}(x_2)\right)\Big{|}_{x_1,x_2\rightarrow x}\,.
\ee  
When acting with $\partial^2$ on the first term, we use the equation of motion (\ref{sigmaeq}). When acting on the second term, on the other hand, 
we can actually use the free equation of motion $\partial^2 \varphi^a = 0$ to this order, so that 
$\partial_1^2 \varphi^{a}\varphi^{a}(x_1) = 2\partial_{\mu}\varphi^a\partial^{\mu}\varphi^a$. The final result for the descendant to 
order $g^2$ can then 
be written in the form
\begin{equation} 
\begin{aligned}
& \partial_{\mu}D^{\mu}_z\hat J_s  = g^2 \hat K_{s-1}\\
&\hat K_{s-1} = -\left(h_s(\deh_1+\deh_3+\deh_4,\deh_2)+h_s(\deh_2+\deh_3+\deh_4,\deh_1)-h_s(\deh_4+\deh_3,\deh_1+\deh_2)\right)\times \\
 & ~~~~~~~~~~~~~~~~~~~~~~~~~~~~~~~~~~~~~~~~~~~\times 
\partial_{3\mu} \partial_{4\mu}\varphi^a(x_1) \varphi^a(x_2) \varphi^b(x_3) \varphi^b(x_4)\Big{|}_{x_{1,2,3,4}\rightarrow x}\\
&h_s(u,v) = 2v(u+v)^{s-2}C_{s-2}^{3/2}(\frac{u-v}{u+v})
\end{aligned}
\end{equation}

Note that all $O(N)$ indices here run from $1$ to $N-1$. The rest of the calculation is almost exactly the same as in the $d=4-\eps$ case. 
Computing the descendant two-point function, using (\ref{mainform}) for $d=2$ and the current two-point function (\ref{JJ-singlet}) (with $N\rightarrow N-1$), 
we find the result
\be
\label{SingleSigma}
\gamma_s=\frac{g_*^4}{4\pi^2}(N-2)\left(\frac{1}{s}-\frac{1}{2}+H_{s-2}\right)=\frac{\eps^2}{N-2}\left(\frac{1}{s}-\frac{1}{2}+H_{s-2}\right)
\ee
where $H_{k}=\sum_{n=1}^{k} 1/n$ is the harmonic number. The $1/N$ expansion of this result precisely matches the expansion of (\ref{zhib}) in $d=2+\eps$.
In the large spin limit, we see the logarithmic behavior (since $H_k \sim \log(k)$ at large $k$)
\begin{equation}
\gamma_s = \frac{\eps^2}{N-2}\left(\log(s)+\gamma-\frac{1}{2}-\frac{1}{2s}+O(1/s^2)\right)\,.
\end{equation}
Also, we note that the leading order in $\gamma_s$ is $\eps^2$, although the leading order anomalous dimension of the $\phi$ field is $\eps$ (\ref{sigmaanom}). 
This may seem to contradict the expected $s\rightarrow \infty$ behavior. The simple resolution of this ``paradox" is suggested by looking 
at the large $N$ result (\ref{LargeNInfSpin}) for the singlet currents, expanded near $d=2+\eps$: 
\be
\begin{gathered}
\gamma_s = \frac{\eps}{N} -\frac{\eps}{N} \frac{\Gamma(3+\eps)}{2(1+\eps)}\frac{1}{s^{\eps}} +...=\frac{\eps^2}{N}\log(s)+...\\
\end{gathered}
\ee 
We see that for the singlets the $2\gamma_{\phi}$ term is canceled by the expansion of the second term, coming from the higher-spin current tower, and the $\log(s)$ is exactly what one gets from expanding the harmonic number $H_{s-2}$. As for the non-singlet operators, 
from (\ref{LargeNInfSpin}) one gets $\gamma_{s(ij)}=\frac{\eps}{N} +O(\frac{1}{s^2})$, and it is evident that the leading order is indeed $2\gamma_{\phi}$ 
as expected \cite{Parisi:1973xn, Callan:1973pu,Alday:2007mf, Komargodski:2012ek,Fitzpatrick:2012yx}. Thus, a finite $N$ calculation of the anomalous dimensions of the non-singlet operators 
should yield a result starting at order $\epsilon$, unlike (\ref{SingleSigma}). 
We leave the more detailed discussion of the non-singlet currents for future work. 

\section{Some $d=3$ estimates}
For the $O(N)$ models with $N\ge 3$, we can combine the information from the $d=4-\epsilon$ and $d=2+\epsilon$ expansions to obtain 
some estimates for the anomalous dimensions of the singlet higher spin currents in $d=3$. The simplest way to do this is to use 
a ``two-sided" Pad\'e approximant. For any given physical quantity assumed to be a continuous function of dimension $d$, 
we can construct the Pad\'e approximant
\begin{equation}
\textrm{Pad\'e}_{[m,n]}(d) = \frac{A_0 + A_1 (4-d) + A_2(4-d)^2 + \ldots + A_m (4-d)^m}{1 + B_1 (4-d) + B_2 (4-d)^2 + \ldots + B_n (4-d)^n}\,,
\label{Pade}
\end{equation}
where the coefficients are fixed by matching the known perturbative expansions in $d=4-\eps$ and $d=2+\eps$. Rather than performing this 
procedure on $\gamma_s(d)$ itself, guided by the expected large spin behavior \cite{Parisi:1973xn, Callan:1973pu,Alday:2007mf, Komargodski:2012ek,Fitzpatrick:2012yx}, we find it more convenient to consider the quantity
\begin{equation}
f_s(d) = \gamma_s(d)-2\gamma_{\phi}(d)
\label{fs}
\end{equation}
From the results (\ref{gamma4e}), (\ref{SingleSigma}), we can obtain the $\epsilon$ expansion of this quantity to order $\epsilon^2$. 
Further information in $d=4-\epsilon$ can be obtained using the result of \cite{Braun:2013tva}, who derived the anomalous dimensions of the 
higher spin operators in the $O(N)$ theory to order $\epsilon^3$.\footnote{For $N=1$, the result is known to order $\epsilon^4$ \cite{Derkachov:1997pf}.} 
For the singlet currents, it reads
\begin{equation}
\gamma_s = \frac{(N+2) \lambda^2 \left(s^2+s-6\right)}{128 \pi ^4 s (s+1)}
-\frac{(N+2)(N+8) \lambda^3 \left(16 s (s+1) H_s+s \left(s^3+2 s^2-39 s-16\right)+12\right)}{4096 \pi ^6 s^2 (s+1)^2}+O(\lambda^4)
\end{equation}
where $H_s$ is the harmonic number. This vanishes at $s=2$, as expected. It is also interesting to check the large spin behavior, which yields (using the known result 
for $\gamma_{\phi}$ to order $\lambda^3$, see e.g. \cite{Kleinert:2001ax})
\begin{equation}
\gamma_s = 2\gamma_{\phi} 
-\frac{(N+2) \left(12 \pi ^2\lambda^2+(N+8)\lambda^3 \left(\log(s)- \gamma -5/2\right)\right)}{256 \pi ^6 s^2}+O(1/s^3)\,.
\label{gammasep3}
\end{equation}
We see that a logarithmic term arises at subleading order in the coupling constant, consistently with general expectations 
\cite{Komargodski:2012ek,Alday:2015ota}. Using the value of the critical coupling \cite{Kleinert:2001ax}
\begin{equation}
\lambda_* =\frac{8 \pi ^2 \eps}{N+8}+ \frac{24 \pi ^2(3N+14) \eps^2}{(N+8)^3}+\ldots 
\end{equation}
we can obtain the $\epsilon$ expansion of $\gamma_s$ around $d=4$ to order $\eps^3$. Further using the $\epsilon$ expansions of 
$\gamma_{\phi}$ near $d=2$ and $d=4$, we can get the function $f(d)$ defined in (\ref{fs}) to the same order
\begin{equation}
\begin{aligned}
&f_s(4-\eps) =-\frac{3 \epsilon^2(N+2)}{s(s+1)(N+8)^2}+O(\epsilon^3)\\
&f_s(2+\eps) = \left(-\frac{\epsilon}{N-2}+\frac{(N-1)\epsilon^2}{(N-2)^2}\right)+\frac{\eps^2}{N-2}(\frac{1}{s}-\frac{1}{2}+H_{s-2}) +O(\epsilon^3)\\
\end{aligned}
\end{equation}
where for simplicity we did not write explicitly the $O(\epsilon^3)$ in $d=4-\eps$, it can be read off from (\ref{gammasep3}). 
This allows to construct Pad\'e approximants (\ref{Pade}) with a maximum value $n+m=6$. Carrying out this procedure for general $N$, 
we find that Pad\'e$_{[3,2]}$ (which only uses $f(4-\eps)$ to order $\eps^2$) and Pad\'e$_{[4,2]}$ appear to give 
the best agreement with the analytic large $N$ result (\ref{zhib}) over the full range $2\le d\le 4$, with Pad\'e$_{[3,2]}$ 
in fact working slightly better. 
Using this approximant, we obtain a $d=3$ estimate for the function $f_s(d)$ in (\ref{fs}). To obtain the anomalous dimensions $\gamma_s$, 
we can then add back the contribution $2\gamma_{\phi}$ using the best available estimates that were collected in Table 2 of \cite{Kos:2013tga} 
for a few low values of $N$. The results of this procedure for $s=4,6,8,10$ and for several values of $N$ are listed in the table below.
\begin{table}[h]
\centering
\begin{tabular}{ccccccc}
\hline
\multicolumn{1}{|c|}{N}         & \multicolumn{1}{c|}{3} & \multicolumn{1}{c|}{4} & \multicolumn{1}{c|}{5} 
& \multicolumn{1}{c|}{6} & \multicolumn{1}{c|}{10}& \multicolumn{1}{c|}{20}\\ \hline
\multicolumn{1}{|c|}{$\gamma_{s=4}~~$ (Pad\'e$_{[3,2]}$)}      & \multicolumn{1}{c|}{0.0261} & \multicolumn{1}{c|}{0.0257} & \multicolumn{1}{c|}{0.0208} 
& \multicolumn{1}{c|}{0.0195} & \multicolumn{1}{c|}{0.0158} & \multicolumn{1}{c|}{0.0082} \\ \hline
\multicolumn{1}{|c|}{$\gamma_{s=6}~~$ (Pad\'e$_{[3,2]}$)}       & \multicolumn{1}{c|}{0.0318} & \multicolumn{1}{c|}{0.0310} & \multicolumn{1}{c|}{0.0258} 
& \multicolumn{1}{c|}{0.0240} & \multicolumn{1}{c|}{0.0191} & \multicolumn{1}{c|}{0.0100} \\ \hline
\multicolumn{1}{|c|}{$\gamma_{s=8}~~$ (Pad\'e$_{[3,2]}$)}      & \multicolumn{1}{c|}{0.0342} & \multicolumn{1}{c|}{0.0332} & \multicolumn{1}{c|}{0.0278} 
& \multicolumn{1}{c|}{0.0259} & \multicolumn{1}{c|}{0.0206} & \multicolumn{1}{c|}{0.0110} \\ \hline
\multicolumn{1}{|c|}{$\gamma_{s=10}~~$ (Pad\'e$_{[3,2]}$)}      & \multicolumn{1}{c|}{0.0353} & \multicolumn{1}{c|}{0.0343} & \multicolumn{1}{c|}{0.0289} 
& \multicolumn{1}{c|}{0.0269} & \multicolumn{1}{c|}{0.0214} & \multicolumn{1}{c|}{0.0115} \\ \hline
\multicolumn{1}{l}{} & \multicolumn{1}{l}{}   & \multicolumn{1}{l}{}   & \multicolumn{1}{l}{}   & \multicolumn{1}{l}{} 
& \multicolumn{1}{l}{}   & \multicolumn{1}{l}{} 
\end{tabular}
\caption{Pad\'e estimates for the anomalous dimensions of the singlet currents with $s=4,6,8$ in the 3d critical $O(N)$ models. The estimates 
are obtained by constructing a ``two-sided" Pad\'e approximant of the function (\ref{fs}) and adding at the end the contribution $2\gamma_{\phi}$ using 
the available results collected in \cite{Kos:2013tga}. For $N=10,20$, the value of $\gamma_{\phi}$ is obtained from the large $N$ result known to
order $1/N^3$ \cite{Vasiliev:1981dg,Vasiliev:1982dc}.} 
\label{tableP}
\end{table}
For comparison, the large $N$ formula (\ref{zhib}) gives for $d=3$
\begin{equation}
\gamma_s=\frac{16(s-2)}{(2s-1)3\pi^2}\frac{1}{N}+O(1/N^2)\,.
\end{equation}
Using this for $N=20$, one would get $\gamma_4=0.0077,\gamma_6=0.0098,\gamma_8=0.0108,\gamma_{10} =0.0114$. The results for $s=4$ given in 
Table \ref{tableP} appear to be consistent with the ones given in \cite{Campostrini:1996vd}. 

For $N=1$ and $N=2$, the nonlinear sigma model result cannot be used since there is no perturbative fixed point in $d=2+\epsilon$ for 
these values of $N$. 
Simple Pad\'e approximants of the $d=4-\eps$ result appear to yield poles in $2< d< 4$ in this case, so we will resort to the unresummed 
$\epsilon$ expansion to obtain some estimates. 
For $N=1$, setting 
$\epsilon=1$ in $f(4-\epsilon)$ expanded to order $\epsilon^3$, and adding back the 3d value of $2\gamma_{\phi}^{N=1}=0.0363$ \cite{ElShowk:2012ht,El-Showk:2014dwa,2010PhRvB..82q4433H,Campostrini:2002cf}, 
we obtain the following $d=3$ estimates
\begin{equation}
\begin{aligned}
&\gamma_{s=4}^{N=1} = 0.0240\,,\qquad \gamma_{s=6}^{N=1} =0.0300 \\
&\gamma_{s=8}^{N=1} = 0.0324\,,\qquad \gamma_{s=10}^{N=1} =0.0336\,.
\label{gammaIsing}
\end{aligned}
\end{equation}
While we do not expect these to be high precision results, 
we observe that they appear to be 
quite close to the estimates derived in \cite{Alday:2015ota}. For the spin 4 operator, \cite{Campostrini:1999at} 
obtained the slightly lower value $\gamma_4 = 0.0208(12)$. For $N=2$, following a similar procedure 
and using $2\gamma_{\phi}^{N=2}=0.0381$ \cite{Campostrini:2006ms}, we obtain 
\begin{equation}
\begin{aligned}
&\gamma_{s=4}^{N=2} =0.0252\,,\qquad \gamma_{s=6}^{N=2} =0.0315 \\
&\gamma_{s=8}^{N=2} =0.0340\,,\qquad \gamma_{s=10}^{N=2} =0.0353\,.
\label{gammaO2}
\end{aligned}
\end{equation}
In all cases, we observe that the anomalous dimensions of the higher spin operators is rather small (similarly to what happens for the anomalous 
dimension of $\phi$). From the results in Table \ref{tableP}, and (\ref{gammaIsing}),(\ref{gammaO2}), we also notice some non-monotonic behavior
as a function of $N$, with a maximum between $N=3$ and $N=4$. A qualitatively similar non-monotonic behavior can be observed in the sphere free energy 
\cite{Giombi:2014xxa,Fei:2015oha} and $C_T$ \cite{Kos:2013tga}. It would be interesting to understand better the origin of this behavior and the 
relation between these quantities. 

\section*{Acknowledgments}

We thank Igor Klebanov, Grisha Tarnopolsky and Sasha Zhiboedov for useful discussions.
The work of SG and VK was supported in part by the US NSF under Grant No.~PHY-1318681. 

\appendix 
\section{Technical details on the computation of the descendant 2-point function}
\label{app-tech}
The calculation of the descendant two-point functions is carried out using the relation of Gegenbeauer polynomials to hypergeometric functions (see appendix of \cite{Belitsky:1998gc}). 
We will illustrate the technique to calculate the function $A_s$ (see eq.~\ref{symKij}) as a function of $s$ in terms of Gegenbauer integrals. 
The calculation of the other structures is similar. 
After introducing the Schwinger parametrization, the $h_s$ function can be written as:
\be
\begin{gathered}
h_s(-2\alpha_1 \hat{x} -2\alpha_3 \hat{x}-2\alpha_4\hat{x},-2\alpha_2 \hat{x}) = (-1)^{s-1} (2\hat{x})^{s-1} (\alpha_1 + \alpha_2+ \alpha_3 + \alpha_4)^{s} \tilde{h}_s(1-2\tilde{\alpha}_2);\\
\tilde{\alpha}_n = \frac{\alpha_n}{\alpha_1+ \alpha_2+ \alpha_3 + \alpha_4}, \qquad \tilde{h}_s(x) = C^{3/2}_{s-1}(x) - 3(1-x) C^{5/2}_{s-2}(x)
\end{gathered}
\ee
and analogously for other arguments. The contraction is then compactly written as:
\begin{multline}
\frac{(-1)^{s-1} (2\hat{x})^{2s-2}}{(4\pi^2)^4} \int^{\infty}_{0}  \prod_{n=1}^{4} d\alpha_n (\sum^{4}_{n=1}\alpha_n)^{2s-2} \exp(-x^2\sum^{4}_{n=1}\alpha_n) (\tilde{h}_s (1-2\tilde{\alpha}_2) + (-1)^s \tilde{h}_s (1-2\tilde{\alpha}_1))^2  \\
\end{multline}
It is convenient to separate the integration other the sum $\sum^{4}_{n=1}\alpha_n$ by introducing a delta function $\int^{\infty}_{0} dp \delta(\sum^{4}_{n=1}\alpha_n- p)$:
\begin{multline}
\label{integral}
\int^{\infty}_{0} dp p^{2s-2+3} \exp(-x^2 p)  \int^{1}_{0}  \prod_{n=1}^{4} d\tilde{\alpha}_n \delta(\sum^{4}_{n=1}\tilde{\alpha}_n- 1)  (\tilde{h}_s (1-2\tilde{\alpha}_2) + (-1)^s\tilde{h}_s (1-2\tilde{\alpha}_1))^2\\
=\frac{(2s+1)!}{(x^2)^{2s+2}} \iint\limits_{0<1-\tilde{\alpha}_1-\tilde{\alpha}_2<1} d\tilde{\alpha}_1 d\tilde{\alpha}_2 (1-\tilde{\alpha}_1-\tilde{\alpha}_2) (\tilde{h}_s (1-2\tilde{\alpha}_2) + (-1)^s\tilde{h}_s (1-2\tilde{\alpha}_1))^2
\end{multline}
The goal now is to calculate the integral of functions $\tilde{h}$. First we study the integral of two $\tilde{h}$ with the same argument:
\be
\int^{1}_{0} d\tilde{\alpha} (1-\tilde{\alpha})^2 (\tilde{h}_s (1-2\tilde{\alpha}))^2 
\ee
The idea now is to employ the Rodrigues formula for the Gegenbauer polynomial:
\be
\label{Rodrigues}
C^{\nu}_{s} (1-2\tilde{\alpha})= \frac{4^s}{s!} \frac{\Gamma(s+\nu)\Gamma(s+2\nu)}{\Gamma(\nu)\Gamma(2s+2\nu)} (\tilde{\alpha}(1-\tilde{\alpha}))^{-\nu+1/2}
\frac{d^{s}}{d\tilde{\alpha}^{s}} (\tilde{\alpha}(1-\tilde{\alpha}))^{s+\nu-1/2}
\ee
We again split the integral into two parts:
\be
\int^{1}_{0} d\tilde{\alpha} (1-\tilde{\alpha})^2\big(C^{3/2}_{s-1}(1-2\tilde{\alpha}) - 3(2\tilde{\alpha}) C^{5/2}_{s-2}(1-2\tilde{\alpha})\big)\tilde{h}_s (1-2\tilde{\alpha}_1)
\ee
We now act on the second $\tilde{h}$ with the $C^{3/2}_{s-1}$ using the Rodrigues formula, by integrating by parts $s-1$ times. The boundary terms vanish thanks to the power of $\tilde{\alpha}(1-\tilde{\alpha})$ under the derivative. The pre-factor $(\tilde{\alpha}(1-\tilde{\alpha}))^{-1}$ combines with $ (1-\tilde{\alpha})^2$ combines to $\frac{1}{\tilde{\alpha}}-1$. Then, since $\tilde{h}_s(1-2\tilde{\alpha})$ is a polynomial of degree $s-1$ in $\tilde{\alpha}$:
\be
\tilde{h}_s(1-2\tilde{\alpha})= \sum^{s-1}_{k=0} c_{k} \tilde{\alpha}^{k},
\ee
 after $s-1$ integrations by parts only two terms survive the differentiation
\be
\begin{gathered}
(-1)^{s-1} \frac{d^{s-1}}{d\tilde{\alpha}^{s-1}} c_{s-1} \tilde{\alpha}^{s-1} =(-1)^{s-1}(s-1)! c_{s-1};\\
(-1)^{s-1} \frac{d^{s-1}}{d\tilde{\alpha}^{s-1}} c_{0} \frac{1}{\tilde{\alpha}} =\frac{(s-1)! }{ \tilde{\alpha}^{s}}
\end{gathered}
\ee
The remaining integrals are now of the beta-function type, for instance,
\be
\begin{gathered}
\int^{1}_{0} d\tilde{\alpha}  c_{s-1} (\tilde{\alpha}(1-\tilde{\alpha}))^{s} \\
\int^{1}_{0} d\tilde{\alpha} c_{0} (1-\tilde{\alpha}))^{s}
\end{gathered}
\ee
This is the main idea of the calculation, the rest is basically collecting all the coefficients and applying the same method to the other integrals which will appear (all of them will be of the same type though). For the sake of reference the coefficients of $\tilde{h}_s(1-2\tilde{\alpha})$ are obtained most easily by using the relation of Gegenbauer polynomials to the hypergeometric function:
\be
C^{\nu}_{s} (1-2x) = \frac{(2\nu)_s}{s!} F\big(-s,s+2\nu;\nu+\frac{1}{2}; x)
\ee
(this is the reason we used $1-2\tilde{\alpha}$ as the argument). After collecting all the factors, the overall answer for the integral in (\ref{integral}) is:
\be
\frac{(s-1)s(s+1)(s+2)}{8(2s+1)}
\ee

\bibliographystyle{ssg}
\bibliography{AnomHS}

\end{document}